\documentclass[12pt]{article} 
\usepackage{graphicx}
\usepackage{epsfig} 
\usepackage{latexsym} 
\usepackage{latexsym}
\usepackage{amssymb} 
\usepackage{amsmath} 
\usepackage{cite}
\def\etc{{\it etc}}

\def\eg{{\it e.g.}}

\def\ie{{\it i.e.}}

\def\Tr{{\rm Tr}}

\DeclareSymbolFont{AMSb}{U}{msb}{m}{n}
\DeclareMathSymbol{\IN}{\mathbin}{AMSb}{"4E}
\DeclareMathSymbol{\IZ}{\mathbin}{AMSb}{"5A}
\DeclareMathSymbol{\IR}{\mathbin}{AMSb}{"52}
\DeclareMathSymbol{\Q}{\mathbin}{AMSb}{"51}
\DeclareMathSymbol{\II}{\mathbin}{AMSb}{"49}
\DeclareMathSymbol{\IC}{\mathbin}{AMSb}{"43}
\DeclareMathSymbol{\IP}{\mathbin}{AMSb}{"50}
\DeclareMathSymbol{\IH}{\mathbin}{AMSb}{"48}
\DeclareMathSymbol\IA{\mathalpha}{AMSb}{"41}
\DeclareMathSymbol\IS{\mathalpha}{AMSb}{"53}

\def\Q{{\cal Q}}

\begin{document}



\begin{flushright}
USC-03-07\\
NSF-KITP-03-105\\
DCPT--03/55
\end{flushright}

\bigskip
\bigskip
\begin{center}
   {\Large \bf Wrapped D--Branes as BPS Monopoles: }

\bigskip

{\Large \bf The Moduli Space Perspective }
   \end{center}

\bigskip
\bigskip
\bigskip

\centerline{\bf Jessica K. Barrett${}^\sharp$, Clifford V.
  Johnson${}^\natural$}

\bigskip
\bigskip
\bigskip

\centerline{\it ${}^\sharp$Centre for Particle Theory} \centerline{\it
  Department of Mathematical Sciences} \centerline{\it University of
  Durham} \centerline{\it Durham, DH1 3LE, U.K.}
\centerline{\small \tt j.k.barrett@durham.ac.uk}


\bigskip
\bigskip

  \centerline{\it ${}^\natural$Department of Physics and Astronomy }
\centerline{\it University of
Southern California}
\centerline{\it Los Angeles, CA 90084-0484, U.S.A.}
\centerline{\small \tt johnson1@usc.edu}
\bigskip
\bigskip


  \centerline{\it ${}^\natural$Kavli Institute for Theoretical Physics}
\centerline{\it University of
California}
\centerline{\it Santa Barbara, CA 93106-4030, U.S.A.}
\centerline{\small \tt johnson1@kitp.ucsb.edu}

\bigskip
\bigskip


\begin{abstract}
  We study the four dimensional effective action of a system of
  D6--branes wrapped on the K3 manifold times a torus, allowing the
  volume of the internal manifolds to remain dynamical. An unwrapped
  brane is at best a Dirac monopole of the dual R--R sector field to
  which it couples. After wrapping, a brane is expected to behave as a
  BPS monopole, where the Higgs vacuum expectation value is set by the
  size of the K3.  We determine the moduli space of an arbitrary
  number of these wrapped branes by introducing a time dependent
  perturbation of the static solution, and expanding the supergravity
  equations of motion to determine the dynamics of this perturbation,
  in the low velocity limit.  The result is the hyper--K\"ahler
  generalisation of the Euclidean Taub--NUT metric presented by
  Gibbons and Manton. We note that our results also pertain to the
  behavior of  bound states of  Kaluza--Klein monopoles and  wrapped
  NS5--branes in the $T^4$ compactified heterotic string.

\end{abstract}
\newpage \baselineskip=18pt \setcounter{footnote}{0}

\section{Introduction and Motivation}
\label{sec:introduction}

D--branes are the smallest sources of Ramond--Ramond (R--R) sector
charge in types~I and~II string theory\cite{Polchinski:1995mt}. A
D$p$--brane has an electric coupling to the $(p+1)$--form
Ramond--Ramond sector potential $C_{(p+1)}$.  By Hodge duality, it
couples magnetically to the potential $C_{(7-p)}$. As magnetic
monopoles in the full ten dimensions, the D--branes may be said to be
monopoles of ``Dirac'' type. They have topology, and a completely
fixed spherically symmetric transverse geometry as ten dimensional
objects in perturbative string theory: They are point--like (on scales
coarser than the string length $\ell_{\rm s}$) in their transverse
dimensions while at stringy scales they are only slightly more
interesting, having an effective ``halo'' of interaction radius of
$\ell_{\rm s}=2\pi\sqrt{\alpha^\prime}$, which is still effectively
spherical.

These simple properties are in contrast to those of a BPS magnetic
monopole\cite{'tHooft:1974qc,Polyakov:1974ek,Bogomolny:1976de,Prasad:1975kr},
obtained from the non--trivial dynamics and topology of spontaneously
broken non--Abelian gauge symmetry. They again have gauge bundle
topology which supplies them with magnetic charge, but they have an
adjustable transverse size set by the physical parameters of the
dynamical symmetry breaking. When endowed with multiples of the basic
unit charge, they also have shapes more interesting than simple
spherical geometry.

Perhaps one of the most interesting and rewarding features of
D--branes is the fact that while they are extremely simple objects
with somewhat boring features (such as those listed in the first
paragraph) in the overall scheme of things, they are readily amenable
to being endowed with interesting properties. This is achieved in a
variety of ways, such as allowing them to interact with other branes
(of the same or different type and dimension) by immersion or
intersection, putting them in background fields, or embedding them in
non--trivial geometry.

In particular, D--branes can behave as BPS monopoles, and therefore
are endowed with properties which are sensitive to the non--Abelian
gauge theory dynamics of the string theory vacuum in which they find
themselves.  This is intriguing for a number of reasons.  BPS
monopoles have a size and shape, both of which are tunable by
adjusting the asymptotic values of the extra scalars which appear in
the theory.  At the same time, supersymmetry is preserved, which
allows for a clean study of the properties of such objects. So we are
in a position to study, with the aid of supersymmetry if we wish, the
dynamics of D--branes which have interesting transverse structure and
size. A whole host of highly instructive examples (various gauge
duals, holographic correspondences, {\it etc.,}) which have been
excavated over the last five years or so have shown that such regimes
---where branes stop being simply point--like in their transverse
directions--- are likely to prove the highly instructive in uncovering
the next levels of understanding of the fundamental physics to which
brane dynamics seem to be guiding us. 

In this paper we would like to further examine the story of how
D--branes behave {\it precisely} like BPS monopoles as a result of
being wrapped on the four dimensional manifold K3. The problem that we
tackle is to take the explicit effective action of the superstring
theory, and the static solutions for the D--branes in this K3--wrapped
situation, and derive the effective Lagrangian for the slow relative
motion of an arbitrary number of such branes in arbitrary positions.
Even though the solution for the fields around the branes (involving
gravity, the dilaton, and various higher rank forms) are very
different from those of a BPS monopole, the result for the effective
Lagrangian should be identical to that of the standard BPS monopole
result presented by Gibbons and Manton\cite{Gibbons:1995yw}. We show
that this is the case explicitly. We notice along the way that the
effective action within which we work is also appropriate to the $T^4$
compactified heterotic string. The reduced action has an elegant
symmetry which takes us from one setup to another. The wrapped
D6--brane under this duality becomes\cite{Johnson:1996bf} a bound
state of an H--monopole (wrapped NS5--brane) and a Kaluza--Klein
monopole, which also is expected to behave as a BPS
monopole\cite{Johnson:2001wm,Krogh:1999qr}. Our result proves that
this is the case for that dual picture also.

\section{BPS Monopoles from wrapping branes}

Realising D--branes as BPS monopoles can be achieved, for example, by
wrapping D--branes on K3, as discussed in ref.\cite{Johnson:1999qt}.
In type~IIA superstring theory, for example, a compactification on K3
yields\footnote{See for example, ref.\cite{Aspinwall:1996mn} for a
  review.} a gauge group of $U(1)^{24}$ in the six non--compact
dimensions, generically. This arises from the R--R three--form
$C_{(3)}$ reduced on the 22 two--cycles of K3, from the R--R one--form
$C_{(1)}$, and from the R--R five--form $C_{(5)}$ wrapped on the
four--cycle that is the entire K3. There is a large family of scalars
in the ${\cal N}=2$ gauge multiplets (counting in four dimensional
units), which have the geometrical interpretation as representing the
volume of the cycles on which these forms are reduced, together with
the flux of the Neveu-Schwarz--Neveu-Schwarz (NS--NS) two form
$B_{(2)}$ through them.  The Abelian gauge group can be enhanced at
special points on the moduli space of these scalar vacuum expectation
values (vevs), corresponding to the vanishing of the
$B$--flux\cite{Aspinwall:1995zi}, and at the same time where the
characteristic length scales of these volumes, $\ell_{\rm cycle}$,
reach the value $\ell_{\rm s}=2\pi\sqrt{\alpha^\prime}$.  In this
special situation, new string theory physics appears. In particular,
the D2-- and D4--branes (which couple electrically to the reduced
forms) are wrapping the cycles, appearing as particles in the
non--compact directions. These particles become massless at the
special points, and are the W--bosons of the enhanced non--Abelian
gauge symmetry.

The BPS monopoles of this pattern of spontaneously broken gauge groups
are constructed as D--branes too\cite{Johnson:2001wm}. In six
dimensions, they must have two extended directions ({\it i.e.}, they
are membranes), and they are formed by wrapping D4--branes on the
two--cycles, or by wrapping a D6--brane on the entire K3. These
cycle--wrapped D--branes carry magnetic charges of the $U(1)$ which
arose from direct reduction on the cycle in question. To couple
magnetically to a $q$--form potential in $D$ dimensions is to couple
electrically to its $(D-2-q)$--form potential arising by Hodge
duality.  In six dimensions therefore, our putative monopoles must
have an electric coupling to $C_{(3)}$.  This is achieved for a
wrapped D4--brane because the flux through the two--cycle induces a
single unit of (negative) D2--brane charge, since there is a
world--volume coupling\cite{Douglas:1998uz,Li:1996pq,Witten:1996im} of
lower rank R--R forms to the Chern characteristic $e^{i{\cal F}}$ of
the mixed ``gauge/two--form bundle'', with field strength two--form
${\cal F}=B+2\pi\alpha^\prime F$. A somewhat different mechanism
achieves the same thing for a wrapped D6--brane.  There is a
world--volume coupling of lower rank R--R forms to the square root of
the Dirac genus ${\hat {\cal A}}(\ell_s^2 R)$, where $R$ is the Ricci
two--form.  Since K3 has non--vanishing Pontryagin class, it induces
$C_{(3)}$--form charge in six
dimensions\cite{Bershadsky:1996sp,Green:1997dd}.

Notice that the facts that D6--branes are magnetic sources of
$C_{(1)}$, and that D4--branes are magnetic sources of $C_{(3)}$ in
{\it ten} dimensions are red herrings, from the point of view of
turning them into BPS monopoles under wrapping. If this were not the
case, one could produce BPS monopoles by wrapping on a torus, $T^4$,
for which in superstring theory (as opposed to heterotic string theory
of course) there are no patterns of spontaneous gauge symmetry
breaking. Instead, it is the collection of subtle features of
D--branes mentioned above which endow the wrapped D--branes with
properties which are intimately related to the geometrical properties
of the cycles upon which they are wrapped. Hence, their size and shape
are controlled by the moduli of the cycles. Correspondingly, as the
moduli of the cycles translate into Higgs vevs in the reduced model,
the branes' properties will be controlled by the Higgs vevs, as should
be the case for BPS monopoles.

The above reasoning is satisfying, but certainly not enough to show
beyond doubt that these wrapped branes are indeed BPS monopoles of the
type we know and love. They could carry the same asymptotic charges,
have the same number of moduli, but behave quite differently in
detail. The next step is to demonstrate that the metric on the space
of moduli is the same as for BPS monopoles\footnote{In the original
  enhan\c con paper\cite{Johnson:1999qt}, the metric on the moduli
  space of a single constituent wrapped brane, moving in the
  background produced by the others, was derived. Our result extends
  that case considerably.}. This is what we will show explicitly in
the rest of the paper, at least in the limit where we do not allow the
objects to approach each other too closely. For closer separations
there are instanton corrections to our computation which we leave for
further study. However, note that supersymmetry ensures that our
moduli space metric is hyper--K\"ahler, and it is believed that
additional assumptions of smoothness and the presence of certain
isometries of the metric completes the asymptotic answer into a unique
non--perturbative result. This is known explicitly for the
two--monopole case, with the complete metric being the Atiyah--Hitchin
manifold\cite{Atiyah:1985fd,Atiyah:1988jp}.

\section{Wrapped Brane Solutions}

We will focus on the case of wrapping D6--branes. We do not have to do
this. However, the case of the wrapping of D4--branes is equivalent to
this discussion by T--duality. There is a large $O(20,4)$ symmetry at
our disposal\cite{Aspinwall:1996mn}, with which we can write the
problem in a variety of ways, and the wrapped D6--brane approach of
ref.\cite{Johnson:1999qt} is the one we choose. It would be very
instructive to make the duality of the situation explicit,
since K3 is an interesting manifold to study, but this is not the goal
of this paper.

The wrapping of the D6--branes on K3, which we assume has a volume
$V$, induces a negative amount of D2--brane
charge\cite{Green:1997dd,Bershadsky:1996sp}. Therefore in the limit
where we have a large number, $N$ of D6--branes present, the metric,
dilaton and Ramond--Ramond potentials which correspond to this
scenario are expected to be\cite{Johnson:1999qt}:
\begin{eqnarray}
ds_{S}^{2} & = & Z_{2}^{-1/2} Z_{6}^{-1/2} \eta_{\mu\nu} dx^{\mu} dx^{\nu} 
+ Z_{2}^{1/2} Z_{6}^{1/2} dx^{i}dx^{i} 
+ V^{1/2} Z_{2}^{1/2} Z_{6}^{-1/2} dS_{\rm K3}^{2}  \nonumber \\
e^{2\Phi} & = & g_s^{2} Z_{2}^{1/2} Z_{6}^{-3/2}  \nonumber \\
C^{(3)} & = & (Z_{2}g_s) ^{-1} dx^{0} \wedge dx^{4} \wedge dx^{5}
\nonumber  \\
\label{eq:C7_soln}
C^{(7)} & = & V (Z_{6}g_s) ^{-1} dx^{0} \wedge dx^{4} \wedge dx^{5} \wedge dx^{6}
\wedge dx^{7} \wedge dx^{8} \wedge dx^{9} \ ,
\end{eqnarray}
where $\mu,\nu = 0,4,5$ are the directions tangent to all branes, and
$i,j=1,2,3$ are the directions transverse to all branes, with
\begin{eqnarray}
Z_2 & =& 1 + {r_2\over r}\ , \qquad r_2  =
-\frac{(2\pi)^4g_sN\alpha'^{5/2}}{2V}\ ,
\nonumber \\
Z_6 & = &1 + \frac{r_6}{r}\ , \qquad r_6  =
\frac{g_sN\alpha'^{1/2}}{2}\ .
\label{charges}
\end{eqnarray}
This solution has $N$ units of D6--brane charge, and consequently $N$
units of negative D2--brane charge.  Notice that as one moves in from
$r=+\infty$ toward smaller $r$, the volume of K3 reduces from $V$.
This is the result of the solution for the wrapped brane
back--reacting on the geometry. In fact, the geometry apparently has a
repulsive
singularity\cite{Behrndt:1995tr,Kallosh:1995yz,Cvetic:1995mx} at $r =
|r_2|$, where the effective K3 volume vanishes.  However, in
ref.\cite{Johnson:1999qt} it was shown that {\it for sufficiently
  large $N$} the geometry given by equations~\eqref{eq:C7_soln} is
only correct down to the ``enhan\c con radius'':
\begin{equation}
r_{\rm e} = \frac{2V}{V - V_*} |r_2|\quad  > \quad |r_2|\ ,
\label{enhanconradius}
\end{equation} where $V_*\equiv \ell_{\rm
  s}^4=(2\pi)^4(\alpha^\prime)^2$, and we assume that $V>V_*$.  In
fact, all the D6--branes live on the sphere at $r = r_{\rm e}$, inside
of which the geometry is flat.

Inside the enhan\c con radius there is an enhanced $SU(2)$ gauge
symmetry (hence the name enhan\c con).  These facts were deduced in
part by appealing to a probe computation: A single D6--brane was used
to probe the background due to all of the others. The brane becomes
unphysical for motion in the background as written, once one proceeds
inside the enhan\c con radius. This is because a constituent brane's
tension would become negative in that region due to the stringy
effects (induced charges) arising from wrapping. The simplest solution
is to declare the geometry in that region to be unphysical, and
replace it by flat space, since there are no sharp sources in the
interior\footnote{This aspect of the construction found further
  support in the supergravity computations presented in
  ref.\cite{Johnson:2001wm}.}. This is not unreasonable, since the
geometry was written down in supergravity with no reference to
possible stringy effects in the first place, being designed to match
only the asymptotic charges.

For a large number, $N$, of D6--branes sourcing the background
geometry, it is consistent to neglect the back--reaction of the probe
on the background and do a probe computation as in
ref.\cite{Johnson:1999qt}, if one assumes that one can freeze the
moduli of all of the other branes. The result was the Euclidean
Taub--NUT metric, with negative mass parameter set by $r_{\rm e}$, and hence
proportional to $N$.

In the computation on which we report here, we unfreeze all of the
moduli. We will put the branes in arbitrary positions, not all clumped
together.  Furthermore we will let the branes explore all of the
moduli available to them, and show that in the limit where we restrict
to slow relative motion, the problem is tractable, and yields a simple
result, the multi--Taub--NUT generalisation written down by Gibbons
and Manton\cite{Gibbons:1995yw}.  In ref.\cite{Johnson:2000bm}, it was
argued that when we arrange all the moduli (except four of them) as in
ref.\cite{Johnson:1999qt}, placing $N-1$ of the branes on a sphere of
radius $r_{\rm e}$, we should expect to recover the moduli space
result derived there, Taub--NUT with negative mass $r_{\rm e}$. This
should be accurate for large $N_a$, a limit where the necessary deviations
away from spherical symmetry for a multi--monopole configuration
should be suppressed.

In deriving the result for the $4N$ dimensional moduli space of BPS
monopoles\cite{Weinberg:1979ma}, our technique follows closely that of
Ferrell and Eardley in refs.\cite{Ferrell:1987gf,Ferrell:thesis}, who
calculated the metric on moduli space for slowly moving
Reissner--Nordstr\"om black holes, using the ideas pioneered by Manton
in ref.\cite{Manton:1982mp}. To make contact with those techniques, we
first dimensionally reduce our action to lower dimensions. We reduce
not just on K3, but on an additional $T^2$ as well, making the
monopoles localised objects. The torus produces no additional
subtleties, since it has no non--trivial structure.  In the next
section we perform the dimensional reduction.

\section{Dimensional Reduction to Four Dimensions}
\label{sec:dimensional_reduction}

In our calculations we will assume that the D6--branes are reasonably
far apart from one another. We dimensionally reduce over the D6--brane
directions, chosen to be $x^4,\ldots,x^9$, to obtain a
four--dimensional picture in which they behave like distinct localised
objects. From the work of ref.\cite{Johnson:2001wm}, we know that if
there are $N_a$ coincident D6--branes at the $a$th position, the lower
limit on its effective size is set by the enhan\c con radius given in
equation~(\ref{enhanconradius}). We should assume that they are
separated by at least that size, but since $N_a$ will be taken to be
unity, the corrections to the enhan\c con radius are significant (it
is no longer a sharp radius, as it is for large $N$ (see discussions
in ref.\cite{Johnson:2000bm}), and so we expect each brane to be
separated from each other by distances significantly greater than
that, to avoid needing to worry about instanton corrections to our
result. The typical separation at which instanton corrections become
significant can be estimated by comparing the Atiyah--Hitchin manifold
to the four--dimensional Euclidean negative mass Taub--NUT solution to
which it reduces for large separations. This is the effective core or
enhan\c con radius for a single D6--brane\cite{Johnson:2000bm}.

\subsection{The Type IIA Action in the Einstein Frame}

We start with the type IIA ten dimensional supergravity action in the string
frame with a D6--brane field strength and a D2--brane field strength excited:
\begin{equation}
\label{eq:action1}
S^{S}_{IIA} = \frac{1}{2 \kappa_{0}^{2}} \int\! d^{10}\! x \,\, \sqrt{-g^{S}}   \,\, \left\{ e^{-2\Phi} [
R^{S} + 4 ( \nabla^{S} \Phi )^{2} ] - \frac{1}{2.4!} ( {F}_{S}^{(4)})^{2} -
\frac{1}{2.8!} ( F_{S}^{(8)})^{2} \right\}\ .
\end{equation}
We will work mostly in the Einstein frame, defined by setting:
\begin{equation}
\label{eq:1}
g^{S}_{\mu\nu} = e^{\frac{(\Phi-\Phi_{0})}{2}}g^{E}_{\mu\nu}\ ,
\end{equation}
where $\Phi_{0}$ is the reference value of the dilaton field $\Phi$,
and $g_s = e^{\Phi_{0}}$ is the string coupling. Let $\tilde{\Phi} =
(\Phi-\Phi_{0})$. Then, after some algebra, the Einstein frame action is:
\begin{eqnarray}
S^{E}_{\rm IIA} 
 =  \frac{1}{2\kappa^2} \int \! d^{10}\! x \,\, \sqrt{-g^{E}} \,\, \bigg\{ 
R^{E} - \frac{1}{2} (\nabla^{E} \tilde{\Phi}) ^{2} - 
\frac{1}{2.4!}e^{\frac{\tilde{\Phi}}{2}} (F_{E}^{(4)}) ^{2} 
 - \frac{1}{2.8!} e^{-\frac{3\tilde{\Phi}}{2}} (F_{E}^{(8)}) ^{2}
 \bigg\}\ ,
\label{eq:action2}
\end{eqnarray}
where $2\kappa^2_0g_s^2=16\pi G_{\rm N}=(2\pi)^7(\alpha^\prime)^4$
sets the ten dimensional Newton's constant $G_{\rm N}$. In what
follows we will relabel $\tilde{\Phi}$ as $\Phi$, and we have absorbed
a factor of $g_s$ into the R--R potentials in going to the Einstein frame.

\subsection{Reduction on K3}
\label{sec:K3_red}

Here we dimensionally reduce on K3. After the dimensional reduction it
will be necessary to perform a conformal transformation so that the
gravity part of the dimensionally reduced action is of the canonical
Einstein--Hilbert form.

We rewrite the action in equation~\eqref{eq:action2} as
\begin{equation}
\label{eq:action3}
S_{\rm IIA} = \frac{1}{2\kappa^2} \int\! d^{10}\! x \,\, \sqrt{-\hat{g}} \,\, \bigg\{ \hat{R} -
\frac{1}{2} (\hat{\nabla} \Phi) ^{2} - 
\frac{1}{2.4!}e^{\frac{\Phi}{2}} (\hat{F}^{(4)}) ^{2} - 
\frac{1}{2.8!} e^{-\frac{3\Phi}{2}} (\hat{F}^{(8)}) ^{2} \bigg\}\ ,
\end{equation}
where we have relabeled all the fields with hats to indicate that they are
10-dimensional fields. The hat in $(\hat{F}^{(4)})^{2}$ also indicates 
that the metric $\hat{g}_{\mu\nu}$ is used to compute the square, {\it
  i.e.,}
\begin{equation}
(\hat{F}^{(4)})^{2} = \hat{g}^{\mu_{1}\nu_{1}} \hat{g}^{\mu_{2}\nu_{2}}
\hat{g}^{\mu_{3}\nu_{3}} \hat{g}^{\mu_{4}\nu_{4}}
\hat{F}^{(4)}_{\mu_{1}\mu_{2}\mu_{3}\mu_{4}}
\hat{F}^{(4)}_{\nu_{1}\nu_{2}\nu_{3}\nu_{4}}\ .
\end{equation}
$\hat{F}^{(4)}$ is a 4-form field strength with potential
$\hat{C}^{(3)} = d\hat{F}^{(4)}$. Similar remarks apply to
$\hat{C}^{(7)} = d\hat{F}^{(8)}$. We wish to calculate the
dimensionally reduced version of equation \eqref{eq:action3} when the
directions $x^{6}$, $x^{7}$, $x^{8}$, $x^{9}$ are compactified on a K3
manifold. We set
\begin{equation}
\hat{g}_{MN} = 
\left( \begin{array}{cc}
\bar{g}_{\mu\nu} & 0 \\
0 & V^{1/2} e^{\beta/2}g^{\rm K3}_{ij}
\end{array} \right)
\end{equation}
where $M,N = 0,\ldots,9$, $\mu,\nu = 0,\ldots,5$, $i,j = 6,\ldots,9$ and
$V_{\rm K3} = V e^{\beta}$ is the volume of the K3 manifold, with $V$ 
constant. 

Since we are assuming that the wrapped branes are localised in the
four dimensions, we can take $\bar{g}_{\mu\nu}$, $\beta$, $\Phi$,
$\hat{F}^{(8)}$ to be independent of the compactified directions
$x^{i}$. We also take the metric on K3, $g^{\rm K3}_{ij}$, to depend only
on the $x^{i}$. Then
\begin{equation}
\label{eq:R1}
\hat{R} = \bar{R} - \frac{5}{4} ( \bar{\nabla} \beta )^{2} -
2 ( \bar{\nabla}^{2} \beta )\ ,
\end{equation}
where we have used that K3 is Ricci flat. Also, we have 
\begin{equation}
\label{eq:det_g}
\sqrt{-\hat{g}} = V e^{\beta} \sqrt{-\bar{g}}\ .
\end{equation}

We make the choice that each non--vanishing component of
$\hat{C}^{(7)}$ contains the indices 6,7,8,9. This is consistent with
the form of the solution given in equation~\eqref{eq:C7_soln}. We set
\begin{equation}
\bar{C}'^{(3)}_{\mu_{1}\mu_{2}\mu_{3}} = V^{-1} \hat{C}^{(7)}
_{\mu_{1}\mu_{2}\mu_{3}6789}\ ,
\end{equation}
and we have used a prime to distinguish the dimensionally reduced D6--brane
potential $\bar{C}'^{(3)}$ from the D2--brane potential, which in the
dimensionally reduced action we will denote $\bar{C}^{(3)}$. Then
\begin{equation}
\label{eq:F8_2}
(\hat{F}^{(8)})^{2} = 8.7.6.5. e^{-2\beta} (\bar{F}'^{(4)})^{2} 
\end{equation}
Also
\begin{equation}
\label{eq:C3}
\hat{C}^{(3)}_{\mu_1\mu_2\mu_3} = \bar{C}^{(3)}_{\mu_1\mu_2\mu_3}\ ,
\end{equation}
since $\hat{C}^{(3)}$ does not have components in the K3 directions.

Substituting \eqref{eq:R1} - \eqref{eq:C3} into the action \eqref{eq:action3},
we get the six dimensional action
\begin{eqnarray}
S_{\rm IIA} & = & \frac{1}{2\kappa^2} \int\! d^{6}\! x \,\, V e^{\beta} \sqrt{-\bar{g}} \,\, 
\bigg[ \bar{R} - 
\frac{5}{4} (\bar{\nabla} \beta) ^{2} - 
2 ( \bar{\nabla} ^{2} \beta ) -
\frac{1}{2} ( \bar{\nabla} \Phi ) ^{2} \nonumber \\
\label{eq:action5}
& & \hskip2cm - \frac{1}{2.4!} e^{\frac{\Phi}{2}} (\bar{F}^{(4)}) ^{2} -
\frac{1}{2.4!} e^{-\frac{3\Phi}{2}} e^{-2\beta} 
(\bar{F}'^{(4)}) ^{2} \bigg]\ .
\end{eqnarray}

This action is not of the standard Einstein form since there is a
factor of $e^{\beta}$ multiplying $\bar{R}$. In order to remove this
factor we perform the following conformal transformation:
\begin{equation}
\label{eq:conf_trans_1}
\bar{g}_{\mu\nu} = e^{-\beta/2} \tilde{g}_{\mu\nu}\ .
\end{equation}
After some algebra, we find that the action is:
\begin{eqnarray}
S_{\rm IIA}  =  \frac{V}{2\kappa^2} \int\! d^{6}\! x \sqrt{-\tilde{g}} \,\,
\bigg[ \tilde{R} - 
\frac{1}{2} (\tilde{\nabla} \beta) ^{2} - 
\frac{1}{2} ( \tilde{\nabla} \Phi ) ^{2}
 - \frac{1}{2.4!} e^{\frac{\Phi}{2}} e^{\frac{3\beta}{2}}
(\tilde{F}^{(4)}) ^{2} -
\frac{1}{2.4!} e^{-\frac{3\Phi}{2}} e^{-\beta/2}
(\tilde{F}'^{(4)}) ^{2} \bigg]\ .\nonumber \\
\label{eq:action6}
\end{eqnarray} 
where we have discarded  total derivative terms. 
  
\subsection{Reduction on $T^{2}$}
\label{sec:T2_red}
Next, we dimensionally reduce the action of
equation~\eqref{eq:action6} on the directions $x^{4},x^{5}$, which
we will compactify on a $T^{2}$.  We set
\begin{equation}
\tilde{g}_{MN} = 
\left( \begin{array}{cc}
\bar{\bar{g}}_{\mu\nu} & 0 \\
0 & e^{\rho} \delta_{ij}
\end{array} \right)
\end{equation}
where $M,N = 0,\ldots,5$, $\mu,\nu = 0,\ldots,3$, $i,j = 4,5$. As in
section \ref{sec:K3_red} we assume that all fields are independent of
the compactified directions $x^{4},x^{5}$. Then:
\begin{equation}
\label{eq:R3}
\tilde{R} = \bar{\bar{R}} - 2 \bar{\bar{\nabla}} ^{2} \rho - 
\frac{3}{2} (\bar{\bar{\nabla}} \rho) ^{2}\ ,
\qquad {\rm and}\qquad
\sqrt{-\tilde{g}} = e^{\rho} \sqrt{-\bar{\bar{g}}}\ .
\end{equation}
We make the choice that the non--vanishing components of
$\tilde{C}^{(3)}$ and $\tilde{C}'^{(3)}$ contain the indices 4,5,
which is consistent with the solutions~\eqref{eq:C7_soln}. We set
\begin{eqnarray}
\bar{\bar{C}}^{(1)}_{\mu} & = & \tilde{C}^{(3)}_{\mu45}\ ,\qquad
\bar{\bar{C}}'^{(1)}_{\mu}  =  \tilde{C}'^{(3)}_{\mu45}\ .
\end{eqnarray}
Then
\begin{eqnarray}
\label{eq:F4_sq3a}
(\tilde{F}^{(4)})^{2}  =  4.3.e^{-2\rho} \bar{\bar{F}}^{2}  \
,\qquad 
(\tilde{F}'^{(4)})^{2} =  4.3.e^{-2\rho} \bar{\bar{F}}'^{2} \ .
\end{eqnarray}
Substituting the above into the action 
\eqref{eq:action6} gives the four-dimensional action:
\begin{eqnarray}
S_{\rm IIA} & = & \frac{L^{2}V}{2\kappa^2} \int\! d^{4}\! x  \,\, e^{\rho} 
\sqrt{-\bar{\bar{g}}} \,\, \bigg[ \bar{\bar{R}} - 
2\bar{\bar{\nabla}} ^{2} \rho - \frac{3}{2} (\bar{\bar{\nabla}} \rho) ^{2} - 
\frac{1}{2} (\bar{\bar{\nabla}} \beta) ^{2} - 
\frac{1}{2} ( \bar{\bar{\nabla}} \Phi ) ^{2} \nonumber \\
\label{eq:action7}
& & \hskip3.5cm- \frac{1}{4} e^{\frac{\Phi}{2}} e^{3\beta/2} e^{-2 \rho}
(\bar{\bar{F}}^{(2)}) ^{2} -
\frac{1}{4} e^{-\frac{3\Phi}{2}} e^{-\beta/2} e^{-2 \rho}
(\bar{\bar{F}}'^{(2)}) ^{2} \bigg] \ ,
\end{eqnarray} 
where $L$ is the length of the compactified dimensions. To restore the
action to canonical form, we make the conformal transformation:
\begin{equation}
\label{eq:conf_trans_2}
\bar{\bar{g}}_{\mu\nu} = e^{-\rho} g_{\mu\nu}\ ,
\end{equation}
and the resulting action is:
 
\begin{eqnarray}
S_{\rm IIA} & = & \frac{L^{2}V}{2\kappa^2} \int\! d^{4}\! x \sqrt{-g} \,\, 
\bigg[ R - ( \nabla \rho) ^{2} - 
\frac{1}{2} (\nabla \beta) ^{2} - 
\frac{1}{2} ( \nabla \Phi ) ^{2} \nonumber \\
\label{eq:action8}
& &\hskip3.5cm - \frac{1}{4} e^{\frac{\Phi}{2}} e^{\frac{3\beta}{2}} e^{- \rho}
(F^{(2)}) ^{2} -
\frac{1}{4} e^{-\frac{3\Phi}{2}} e^{-\beta/2} e^{- \rho}
(F'^{(2)}) ^{2} \bigg]\ ,
\end{eqnarray} 
where we have dropped all adornments on the four dimensional
quantities, for clarity.

\subsection{A Symmetry of the Action and the Dual Heterotic String}

Note that the six dimensional action \eqref{eq:action6} is invariant
under the transformation
\begin{equation}
\Phi \longleftrightarrow -\beta\ ,
\quad\mbox{and}\quad F^{(4)} \longleftrightarrow F'^{(4)}\ ,
  \label{dualityone}  
\end{equation}
and the four dimensional action is
invariant under
\begin{equation}
\Phi \longleftrightarrow -\beta\ ,
\quad\mbox{and}\quad F^{(2)} \longleftrightarrow
F'^{(2)}
  \label{dualitythree}\ .
\end{equation}
That is, the dilaton is exchanged with the volume of K3, while the
D2--brane and D6--brane potentials are interchanged.

This symmetry is a consequence of, or consistent with (depending upon
taste), the duality between Type IIA strings compactified on K3 and
heterotic strings compactified on $T^4$. Under this duality, the
dilaton field $e^{\Phi}$, which plays the role of the Type IIA string
coupling in the Type IIA string action, becomes the volume of the
$T^4$ in the heterotic string action. Conversely, the field
$e^{\beta}$ plays the role of the volume of K3 in the Type IIA action,
and the role of the heterotic string coupling in the heterotic string
action.

In terms of the field strengths, the $F^{(8)}$ field strength in the
ten-dimensional Type IIA action is Hodge dual to a $F^{(2)}$ field
strength.  The $F^{(2)}$ field strength is not wrapped in the
six--dimensional Type IIA theory. Under the heterotic--Type IIA duality
this $F^{(2)}$ field strength becomes an $F^{(2)}$ field strength in
the heterotic theory, which is wrapped on the $T^4$ directions.
Therefore in the ten--dimensional heterotic string theory the
corresponding field strength is $F^{(6)}$, which is Hodge dual to
$F^{(4)}$. So the $F^{(8)}$ in the ten dimensional string theory
becomes $F^{(4)}$ in the heterotic theory, and vice versa, as the
transformation requires. In other words, the D6--brane charges in the
Type IIA string theory are transformed\cite{Johnson:1996bf} into
Kaluza--Klein monopole charges in the heterotic string theory, and the
D2--brane charges are transformed into H--monopole (wrapped
NS5--brane) charges. Our results in this paper therefore pertain to
the bound state of these two objects in the heterotic theory, which
also is expected to behave as a BPS
monopole\cite{Johnson:2001wm,Krogh:1999qr}.

\subsection{The Static Solution}
\label{sec:static-solution}

We take as our static solution the form we displayed in equations
(\ref{eq:C7_soln}), with the assumption being that
$Z_{2}$ and $Z_{6}$ are general harmonic functions of the $x^{i}$. We
convert the string frame solution to the Einstein frame using
\begin{eqnarray}
g_{\mu\nu}^{E} & = & e^{-(\Phi-\Phi_{0})/2} g_{\mu\nu}^{S} 
 =  Z_{2}^{-1/8}Z_{6}^{3/8} g_{\mu\nu}^{S}  \nonumber \ .
\end{eqnarray}
Then the ten dimensional solution in the Einstein frame is
\begin{eqnarray}
ds^2 & = & Z_{2}^{-5/8} Z_{6}^{-1/8} \eta_{\mu\nu} dx^{\mu} 
dx^{\nu} 
+ Z_{2}^{3/8} Z_{6}^{7/8} dx^{i}dx^{i} 
  + V^{1/2} Z_{2}^{3/8} Z_{6}^{-1/8} dS_{\rm K3}^{2}\ , \nonumber \\
e^{2\Phi} & = & Z_{2}^{1/2} Z_{6}^{-3/2}\ , \nonumber \\
\nonumber C^{(3)} & = & (Z_{2}) ^{-1} dx^{0} \wedge dx^{4} \wedge
dx^{5} \ ,\\
\label{eq:soln1_4}
C^{(7)} & = & V (Z_{6}) ^{-1} dx^{0} \wedge dx^{4} \wedge dx^{5} \wedge dx^{6}
\wedge dx^{7} \wedge dx^{8} \wedge dx^{9}\ ,   
\end{eqnarray}
where again we have relabeled $\tilde{\Phi}$ as $\Phi$.  We wish to
compactify this ten dimensional solution following the same steps as
in sections \ref{sec:K3_red} and \ref{sec:T2_red} to obtain the four
dimensional solution to the action \eqref{eq:action8}.
After some computation, the solution to the action \eqref{eq:action6} is
\begin{eqnarray}
\widetilde{ds}^{2} & = & Z_{2}^{-1/4} Z_{6}^{-1/4} \eta_{\mu\nu} dx^{\mu} dx^{\nu} 
+ Z_{2}^{3/4} Z_{6}^{3/4} dx^{i}dx^{i}\ ,  \nonumber \\ 
e^{\beta} & = & Z_{2}^{3/4} Z_{6}^{-1/4}\ , \qquad
e^{2\Phi}  =  Z_{2}^{1/2} Z_{6}^{-3/2} \nonumber  \\
\tilde{C}^{(3)} & = & (Z_{2}) ^{-1} dx^{0} \wedge dx^{4} \wedge
dx^{5}\ ,
\nonumber  \\
\label{eq:soln2_5}
\tilde{C}'^{(3)} & = & (Z_{6}) ^{-1} dx^{0} \wedge dx^{4} \wedge
dx^{5} \ .
\end{eqnarray} 

After compactifying on $T^2$, some computation gives the four
dimensional solution  to the action \eqref{eq:action8} as:
\begin{eqnarray}
ds^{2} & = & -Z_{2}^{-1/2} Z_{6}^{-1/2} dx^{0} dx^{0} 
+ Z_{2}^{1/2} Z_{6}^{1/2} dx^{i}dx^{i}  \nonumber\\
e^{\rho} & = & Z_{2}^{-1/4} Z_{6}^{-1/4}\ ,\qquad
e^{\beta}  =  Z_{2}^{3/4} Z_{6}^{-1/4}\ , \qquad 
e^{2\Phi}  =  Z_{2}^{1/2} Z_{6}^{-3/2} \nonumber  \\
C^{(1)} & = & (Z_{2}) ^{-1} dx^{0}\ ,\qquad C'^{(1)}  =  (Z_{6}) ^{-1} dx^{0} \ .
\label{eq:soln3_6}
\end{eqnarray}

\section{The Multi--Wrapped--Brane Moduli Space}

We observe that the dimensionally reduced action in
equation~\eqref{eq:action8} is identical to the action for
four-dimensional gravity, with two $U(1)$ gauge potentials, coupled to
three scalar fields, as we would expect. Moreover, the form of the
four--dimensional enhan\c con solution~\eqref{eq:soln3_6} is very
similar to that of Reissner--Nordstr\"om black holes in the
Einstein--Maxwell--dilaton system, which is given by:
\begin{eqnarray}
\label{eq:RN_soln1}
ds^2 & = & -U^{-2}(\vec{x}) dt^2 + U^2(\vec{x}) d\vec{x}^2 \nonumber \\
e^{-2\Phi} & = & F(\vec{x})\ ,\quad
A  =  (F(\vec{x}))^{-1} dt\ ,
\end{eqnarray}
with
\begin{equation}
U(\vec{x})  =  (F(\vec{x}))^{1/2}  \quad \mbox{and} \quad
F(\vec{x})  =  1 + \sum_a \frac{\mu_a}{|\vec{x} - \vec{x}_a|}  
\label{eq:RN_soln5}
\end{equation}
where $\vec{x}_a$ is the position of the $a$th black hole, and $\mu_a$ is its
mass, and $A$ is the  $U(1)$ potential.

The metric on moduli space for Reissner--Nordstr\"om black holes in
the Einstein--Maxwell system was calculated by Ferrell and Eardley in
ref.\cite{Ferrell:1987gf}, (see also ref.\cite{Gibbons:1986cp}) using
Manton's technique for slowly moving solitons, outlined in
\cite{Manton:1982mp}. This work was extended to the
Einstein--Maxwell--dilaton system by Shiraishi in
\cite{Shiraishi:1993ia}, and first applied to the study of strings in
ref.\cite{Callan:1996hn} and to branes in ref.\cite{Khuri:1996ae}.
Here we will follow the procedure of the Ferrell and Eardley papers,
this time for a system of $N$ K3--wrapped D6--branes.

\subsection{The Static Solution}

We start with the static four--dimensional solution given by
equations~\eqref{eq:soln3_6}, where we will henceforth relabel
$C'^{(1)}$ as $\tilde{C}^{(1)}$. The solutions for the metric,
$e^{\Phi}$, $C^{(1)}$ and $\tilde{C}^{(1)}$ are in agreement with the
Reissner--Nordstr\"om solution given in equations~\eqref{eq:RN_soln1}
and~\eqref{eq:RN_soln5} if we take $U = Z_2^{1/4}Z_6^{1/4}$ and $Z_2 =
Z_6$. The main schematic difference between the black hole solution
and our solution is the extra scalar fields $\beta$ and $\rho$. The
results for the low energy scattering will be very different however,
as we shall see. 

\subsection{Point Sources and a Regulator}
\label{sec:point-sourc-regul}
Since we are taking the limit where the branes are kept a reasonable
distance apart, and since we are in the supergravity approximation
with a small number of branes at each core, we will be able to
legitimately treat them as point--like. Therefore the source terms for
the $U(1)$ charges in the action have the form of $\delta$--functions.
Then the equations of motion for $C^{(1)}$ and $\tilde{C}^{(1)}$ imply
that $Z_2$ and $Z_6$ obey the equations:
\begin{eqnarray}
\nabla^2 Z_2 & = & \sum_{a=1}^N (r_2)_a \delta^{(3)} ( \vec{x} - \vec{x}_a)\
,\quad {\rm and}\quad
 \label{eq:Z6_eqn}
\nabla^2 Z_6  =  \sum_{a=1}^N (r_6)_a \delta^{(3)} ( \vec{x} - \vec{x}_a)\ ,
\end{eqnarray}
where the positions of the branes are denoted $\vec{x}_a$. The
$\vec{x}_a$ are the modular parameters of the solution.  Also, we have
\begin{eqnarray}
(r_2)_a  =  -\frac{(2\pi)^4 g_s Q_a \alpha'^{5/2}}{2V}\ ,\qquad
(r_6)_a  =  \frac{g_s Q_a \alpha'^{1/2}}{2}\ , \nonumber
\end{eqnarray}
where $Q_a$ is the number of D6--branes at the $a$th position. We will
ultimately set this number to unity.  Equations \eqref{eq:Z6_eqn} have
the solutions:
\begin{eqnarray}
Z_2 & = & 1+\sum_{a=1}^N\frac{(r_2)_a}{|\vec{x} - \vec{x}_a|}\ , \quad{\rm and}\quad
\label{eq:Z6_soln}
Z_6  =  1+\sum_{a=1}^N\frac{(r_6)_a}{|\vec{x} - \vec{x}_a|}\ .
\end{eqnarray}
As in ref.\cite{Ferrell:1987gf}, the form of these functions will
produce infinities at various points in the computations. These
divergences hide a lot of interesting physics, in fact, and must be
regularised. We regularise by assuming a general charge
density~$\tilde{Q}$, in effect smearing out the branes. Then
equations~\eqref{eq:Z6_eqn} become:
\begin{eqnarray}
\nabla^2 Z_2 & = & -\frac{(2\pi)^4 g_s \alpha'^{5/2}}{2V} \tilde{Q}(\vec{x})
\quad {\rm and}\quad
\nabla^2 Z_6  =  \frac{g_s \alpha'^{1/2}}{2} \tilde{Q}(\vec{x})\ ,
\end{eqnarray}
and we will take the limit 
\begin{equation}
\tilde{Q} \to \sum_{a=1}^N Q_a \delta^{(3)} (
\vec{x} - \vec{x}_a)
  \label{eq:limit}  
\end{equation}
in the final stages of the calculation, as in
ref.~\cite{Ferrell:1987gf}. Later on, it will be clear that this
scheme actually corresponds to regularising correctly in order to
extract what is in effect, in the
dual\cite{Chalmers:1997xh,Hanany:1997ie} 2+1 dimensional $U(N)$ gauge
theory, the one--loop quantum corrections to the classical physics. In
the monopole moduli space picture, the classical physics is simply a
flat moduli space, and the deviations from this to give a non--trivial
metric is what the regularisation scheme is able to capture in a
controlled manner.

\subsection{Perturbing the Static Solution}
\label{sec:pert-stat-solut}
In the low--velocity approximation we can make the static solutions
time dependent by allowing the moduli to depend on time $\vec{x}_a \to
\vec{x}_a(t)$. We define $\vec{u}_a$ to be the velocity of the $a$th
centre, so that $\vec{u}_a = \vec{\dot{x}}_a(t)$. For the general
charge density we define $\vec{u} = \vec{\dot{x}}(t)$ to be the
velocity of a charged particle of dust.

We perturb the solution~\eqref{eq:soln3_6} (recall we have relabeled
$C'^{(1)}$ as $\tilde{C}^{(1)}$) to take into account the effects of
the time dependence. Since we are assuming that $u = |\vec{u}|$ is
small we only need calculate the perturbed fields to linear order. As
in ref.\cite{Ferrell:1987gf}, first--order perturbations in quantities
which are even under time reversal vanish. Therefore the perturbed
solution can be written in the form (we perform a simple gauge
transformation on the R--R potentials for later convenience):
\begin{eqnarray}
ds^2 & = & -Z_2^{-1/2} Z_6^{-1/2} dt^2 + Z_2^{1/2} Z_6^{1/2} d\vec{x}^2 +
2\vec{N} \centerdot d\vec{x}dt \ ,\nonumber \\
C^{(1)} & = & (1 - (Z_2)^{-1})dt + \vec{A} \centerdot d\vec{x}\ , \nonumber
\\
\label{eq:pert_soln3}
\tilde{C}^{(1)} & = & (1 - (Z_6)^{-1})dt + \vec{\tilde{A}} \centerdot 
d\vec{x} \ ,
\end{eqnarray}
and the scalar fields $\Phi$, $\beta$ and $\rho$ remain
unperturbed. The
perturbations $\vec{A}$, $\vec{\tilde{A}}$ and $\vec{N}$ depend on time through $\vec{x}(t)$.

According to standard lore\cite{Manton:1982mp}, we can neglect
radiation effects, because these effects are of higher order than
$u^2$. Therefore we can assume that the energy in the system remains
in the zero modes; the non--zero modes are not excited. This means that
the motion takes the form of a geodesic in moduli space.

\subsection{The Action in the Slow Motion Limit}
\label{sec:action-slow-motion}
We wish to find the equations of motion for the perturbations
$\vec{N}$, $\vec{A}$ and $\vec{\tilde{A}}$ in order to express these
fields as functions of $\tilde{Q}$ and $u$. We need expressions for
$\vec{N}$, $\vec{A}$ and $\vec{\tilde{A}}$ to $O(u)$, so we must
calculate the perturbed action to $O(u^2)$. Therefore we substitute
the perturbed solutions~\eqref{eq:pert_soln3} into the action,
neglecting terms $O(u^3)$, whence we derive equations of motion from the
resulting approximate action.

In section \ref{sec:dimensional_reduction} we found that the
ten dimensional IIA supergravity action with four dimensions
compactified on K3, and two dimensions compactified on $T^2$ reduces
to the following four--dimensional action
\begin{equation}
\label{eq:fullIIA_action}
S_{\rm IIA} = S_{\rm gravity} + S_{\rm Maxwell} + S_{\rm scalar}\ ,
\end{equation}
where
\begin{eqnarray}
S_{\rm gravity} & = & k \int \! d^4\! x \,\,\sqrt{-g} \,\, R \ ,\nonumber \\
S_{\rm Maxwell} & = & k \int \! d^4\! x \,\, \sqrt{-g} \,\, \bigg( -\frac{1}{4} e^{\Phi/2}
e^{3\beta/2} e^{-\rho} (F^{(2)})^2 - \frac{1}{4} e^{-3\Phi/2} e^{-\beta/2}
e^{-\rho} (\tilde{F}^{(2)})^2 \bigg)\ , \nonumber \\
S_{\rm scalar} & = & k \int d^4x \sqrt{-g} \bigg( -(\nabla \rho)^2 - \frac{1}{2}
(\nabla \beta)^2 - \frac{1}{2}(\nabla \Phi)^2 \bigg)\ ,
\end{eqnarray}
where we have defined the useful  constant, $k$, as
\begin{equation}
  \label{eq:constantk}
   k = L^2V/2\kappa^2\ .
\end{equation}

Substituting the perturbed solutions~\eqref{eq:pert_soln3} into the
action \eqref{eq:fullIIA_action}, and integrating by parts several times,
we find:
\begin{eqnarray}
S_{\rm IIA}^{\rm approx} & = & k \int\! d^4\! x \,\, \bigg\{ 
- \frac{1}{2} \frac{| \vec{\nabla} \times (\vec{A} + Z_2^{-1/2} Z_6^{1/2} 
\vec{N} |^2} {Z_2^{-1} Z_6} 
- \frac{1}{2} \frac{| \vec{\nabla} \times (\vec{\tilde{A}} + Z_2^{1/2} 
Z_6^{-1/2} \vec{N} |^2} {Z_2 Z_6^{-1}}
\nonumber \\
&&\hskip1cm +\frac{ (\vec{\nabla} \times (\vec{A} + Z_2^{-1/2} Z_6^{1/2} \vec{N}))
\centerdot (\vec{\nabla} \times (Z_2^{1/2} Z_6^{1/2} \vec{N} ))}{Z_6}
\nonumber \\
&&\hskip1cm+\frac{ (\vec{\nabla} \times (\vec{\tilde{A}} + Z_2^{1/2} Z_6^{-1/2} \vec{N}))
\centerdot (\vec{\nabla} \times (Z_2^{1/2} Z_6^{1/2} \vec{N} ))}{Z_2}
- \frac{1}{2} \frac{|\vec{\nabla} \times (Z_2^{1/2} Z_6^{1/2} \vec{N}
)|^2}{Z_2 Z_6}
\nonumber \\
\label{eq:IIA_approx_action}
&& \hskip1cm - \dot{Z}_2 \dot{Z}_6 
- \vec{\nabla} (\dot{Z}_2) \centerdot \left( \vec{A} + Z_2^{-1/2} Z_6^{1/2}
\vec{N} \right)
- \vec{\nabla} (\dot{Z}_6) \centerdot \left( \vec{\tilde{A}} + Z_2^{1/2} 
Z_6^{-1/2} \vec{N} \right)
\bigg\}
\end{eqnarray}

We also need to include source terms in the action for the matter
density and for the current. To find the matter source terms we need
to dimensionally reduce the Born--Infeld action for the D6--branes and
the Born--Infeld action for the D2--branes. These are given by:
\begin{equation}
\label{eq:S_matter1}
S_{\rm matter} = - \int\! d^7 \xi\,\, e^{-\Phi} T_6 \sqrt{- \det{\hat{G}^{S}}} 
+ \int\! d^3 \xi\,\, e^{-\Phi} T_2 \sqrt{- \det{\bar{G}^{S}}} \ ,
\end{equation}
where $\xi^\alpha$ are world--volume coordinates. Also
$\hat{G}^{S}_{\alpha\gamma}$ and $\bar{G}^S_{\alpha\gamma}$ are the
induced or ``pulled--back'' metrics on the D6--brane world--volume and
the D2--brane world--volume respectively, \eg\ 
\begin{equation}
{\tilde G}_{\alpha\gamma}=G_{\mu\nu}\frac{\partial X^\mu}{\partial
  \xi^\alpha}\frac{\partial X^\nu}{\partial \xi^\gamma} \ , 
\label{eq:pullitback}
\end{equation}
and $T_6$ and $-T_2$ are
the D6--brane tension and the (negative) D2--brane tension
respectively. We follow the same steps as in section
\ref{sec:dimensional_reduction} to reduce the ten dimensional action
\eqref{eq:S_matter1} to a four-dimensional one; we convert to the
Einstein frame, then compactify on K3$\times T^2$ to get
\begin{equation}
\label{eq:S_matter2}
S_{\rm matter} = - L^2 \int\! dt\,\, e^{-\Phi/4} e^{-3\beta/4} e^{\rho/2}
(e^{\Phi} e^{\beta} V \tau_6 - \tau_2) \sqrt{-\det{G}} 
\end{equation}
where $\tau_p = T_p g^{-1}_s$ is the physical tension for $p=2,6$. Also,
$G_{\alpha\gamma}$ is the metric induced from the four--dimensional metric
in equation~\eqref{eq:pert_soln3}. Substituting the rest of the
perturbed solutions of equations~\eqref{eq:pert_soln3} into the action
\eqref{eq:S_matter2}, we find
\begin{equation}
S^{\rm approx}_{\rm matter} = -L^2 \int\! dt\,\, Z_2^{-1} (VZ_2Z_6^{-1} \tau_6- \tau_2) 
\left( 1 - Z_2^{1/2} Z_6^{1/2} \vec{N} \centerdot \vec{u} - \frac{1}{2} Z_2 Z_6
\vec{u}^2 \right)
\end{equation}
The BPS bounds give $\tau_6 = q_6\mu_6g_s^{-1}$, and 
$\tau_2 = q_2\mu_2g_s^{-1}$, where $q_6$ is the D6--brane charge and $q_2$ is
the D2--brane charge and $\mu_2 = (2\pi)^{-2} \alpha'^{-3/2}$ and $\mu_6 = 
(2\pi)^{-6} \alpha'^{-7/2}$. In terms of the current density
$\tilde{Q}(\vec{x})$ we have
\begin{equation}
q_6 = -q_2 = \int\! d^3\!x \,\, \tilde{Q}(\vec{x})
\end{equation}
So we may write a four dimensional action:
\begin{equation}
\label{eq:S_matter3}
S^{\rm approx}_{\rm matter} = -L^2 \int\! d^4\!x \,\, \frac{\tilde{Q}}{g_s}
\,\, (VZ_6^{-1} \mu_6- Z_2^{-1}\mu_2) 
\left( 1 - Z_2^{1/2} Z_6^{1/2} \vec{N} \centerdot \vec{u} - \frac{1}{2} Z_2 Z_6
\vec{u}^2 \right)\ ,
\end{equation}
The source terms in the action for $C^{(3)}$ and $C^{(7)}$ are given
by the integrated pulls--back:
\begin{equation}
\label{eq:S_current}
S_{\rm current}^{(3)} = \frac{\mu_2q_2}{g_s} \int \,\,  C^{(3)}\
,\quad {\rm and} \quad
S^{(7)}_{\rm current} = \frac{\mu_6 q_6}{g_s} \int\! \,\, C^{(7)}\ .
\end{equation}
Compactifying the actions in \eqref{eq:S_current} on $T^2$ and on $T^2
\times \rm K3$, respectively, and substituting in the perturbed
solution~\eqref{eq:pert_soln3} gives
\begin{eqnarray}
S_{\rm current}^{\rm approx} & = & L^2\int \! dt \left( (Z_2^{-1}-1) + \vec{A} \centerdot
\vec{u} \right)  \frac{q_2 \mu_2}{g_s}
+ L^2\int\! dt \left( (Z_6^{-1}-1) + \vec{\tilde{A}} \centerdot \vec{u}
\right) V \frac{q_6 \mu_6}{g_s} \nonumber \\
\label{eq:S_current3}
& = & L^2\int\! d^4\! x \left( (Z_2^{-1}-1) + \vec{A} \centerdot
\vec{u} \right) \frac{\tilde{Q} \mu_2}{g_s} 
+ L^2\int\! d^4\! x \left( (Z_6^{-1}-1) + \vec{\tilde{A}} \centerdot \vec{u}
\right) \frac{\tilde{Q} V \mu_6}{g_s} \ .
\end{eqnarray}

Altogether we have
\begin{equation}
\label{eq:approx_action}
S^{\rm approx} = S_{\rm IIA}^{\rm approx} + S_{\rm matter}^{\rm approx} +
S_{\rm current}^{\rm approx}\ .
\end{equation}
Substituting equations~\eqref{eq:IIA_approx_action}, \eqref{eq:S_matter3} and 
\eqref{eq:S_current3} into \eqref{eq:approx_action} we get the expression
\begin{eqnarray}
S^{\rm approx} & = & k \int\! d^4\! x \,\, \bigg\{ 
- \frac{1}{2} \frac{| \vec{\nabla} \times (\vec{A} + Z_2^{-1/2} Z_6^{1/2} 
\vec{N} |^2} {Z_2^{-1} Z_6} 
- \frac{1}{2} \frac{| \vec{\nabla} \times (\vec{\tilde{A}} + Z_2^{1/2} 
Z_6^{-1/2} \vec{N} |^2} {Z_2 Z_6^{-1}}
\nonumber \\
&&+\frac{ (\vec{\nabla} \times (\vec{A} + Z_2^{-1/2} Z_6^{1/2} \vec{N}))
\centerdot (\vec{\nabla} \times (Z_2^{1/2} Z_6^{1/2} \vec{N} ))}{Z_6}
\nonumber \\
&&+\frac{ (\vec{\nabla} \times (\vec{\tilde{A}} +
 Z_2^{1/2} Z_6^{-1/2} \vec{N}))
\centerdot (\vec{\nabla} \times (Z_2^{1/2} Z_6^{1/2} \vec{N} ))}{Z_2}
- \frac{1}{2} \frac{|\vec{\nabla} \times (Z_2^{1/2} Z_6^{1/2} \vec{N}
)|^2}{Z_2 Z_6}\ 
\nonumber \\
&& - \dot{Z}_2 \dot{Z}_6 
- \left\{\tilde{Q} \bigg(\frac{1}{g_s}\mu_6V + \frac{1}{g_s}\mu_2 \bigg)
+ \frac{1}{2} \tilde{Q} \left( \frac{V\mu_6Z_2}{g_s} - \frac{\mu_2Z_6}{g_s} \right)
u^2\right\}\frac{L^2}{k} 
\nonumber \\
&& - \left( \vec{A} + Z_2^{-1/2} Z_6^{1/2} \vec{N} \right)  \centerdot
\left( \vec{\nabla} (\dot{Z}_2) + \tilde{Q}\frac{ L^2\mu_2}{g_sk}
  \vec{u}\right)\ 
\nonumber \\
\label{eq:approx_action1}
&&-  \left( \vec{\tilde{A}} + Z_2^{1/2} Z_6^{-1/2} \vec{N} \right) \centerdot
\left( \vec{\nabla} (\dot{Z}_6) - \tilde{Q} \frac{L^2 V \mu_6}{g_sk} \vec{u}\right)
\bigg\}\ .
\end{eqnarray}

\subsection{Perturbation Equations of Motion}
\label{sec:pert-equat-moti}
Since we have calculated $S^{\rm approx}$ up to $O(u^2)$, we can
derive equations of motion from it which are correct to $O(u)$. The
equations of motion for $\vec{A}$ and $\vec{\tilde{A}}$ are
\begin{eqnarray}
- \vec{\nabla} \times \left( \frac{\vec{\nabla} \times (\vec{A} + Z_2^{-1/2}
Z_6^{1/2} \vec{N})} {Z_2^{-1} Z_6} \right) 
+  \vec{\nabla} \times \left( \frac{\vec{\nabla} \times (Z_2^{1/2} Z_6^{1/2}
\vec{N})} {Z_6} \right)
-  \vec{\nabla} \dot{Z}_2 - \frac{\tilde{Q} \mu_2}{g_sk} L^2 \vec{u} & =
& 0 \ ,\nonumber
 \\\label{eq:A_eqn}
&&\\
- \vec{\nabla} \times \left( \frac{\vec{\nabla} \times (\vec{\tilde{A}} + 
Z_2^{1/2} Z_6^{-1/2} \vec{N})} {Z_2 Z_6^{-1}} \right) 
+  \vec{\nabla} \times \left( \frac{\vec{\nabla} \times (Z_2^{1/2} Z_6^{1/2}
\vec{N})} {Z_2} \right)
-  \vec{\nabla} \dot{Z}_6 - \frac{\tilde{Q} V \mu_6}{g_sk} L^2 \vec{u} &
= & 0 \ ,\nonumber\\
&&\label{eq:A_tilde_eqn}
\end{eqnarray}
and the equation of motion for $\vec{N}$ is
\begin{eqnarray}
&- Z_2^{-1/2} Z_6^{1/2} \vec{\nabla} \times \left( \frac{\vec{\nabla} \times 
(\vec{A} + Z_2^{-1/2} Z_6^{1/2} \vec{N})} {Z_2^{-1} Z_6} \right) 
- Z_2^{1/2} Z_6^{-1/2} \vec{\nabla} \times \left( \frac{\vec{\nabla} \times 
(\vec{\tilde{A}} + Z_2^{1/2} Z_6^{-1/2} \vec{N})} {Z_2 Z_6^{-1}} \right) 
\nonumber \\
&+  Z_2^{-1/2} Z_6^{1/2} \vec{\nabla} \times \left( \frac{\vec{\nabla} \times 
(Z_2^{1/2} Z_6^{1/2} \vec{N})} {Z_6} \right)
+ Z_2^{1/2} Z_6^{1/2} \vec{\nabla} \times \left( \frac{\vec{\nabla} \times 
(\vec{A} + Z_2^{-1/2} Z_6^{1/2} \vec{N})} {Z_6} \right)
\nonumber \\
&+  Z_2^{1/2} Z_6^{-1/2} \vec{\nabla} \times \left( \frac{\vec{\nabla} \times 
(Z_2^{1/2} Z_6^{1/2} \vec{N})} {Z_2} \right)
+ Z_2^{1/2} Z_6^{1/2} \vec{\nabla} \times \left( \frac{\vec{\nabla} \times 
(\vec{\tilde{A}} + Z_2^{1/2} Z_6^{-1/2} \vec{N})} {Z_2} \right)
\nonumber \\
&-  Z_2^{1/2} Z_6^{1/2} \vec{\nabla} \times \left( \frac{\vec{\nabla} \times 
(Z_2^{1/2} Z_6^{1/2} \vec{N})} {Z_2 Z_6} \right)
- Z_2^{-1/2} Z_6^{1/2} \left(  \vec{\nabla} \dot{Z}_2 
+ \frac{\tilde{Q} \mu_2}{g_sk} L^2 \vec{u} \right) 
\nonumber \\
&- Z_2^{1/2} Z_6^{-1/2} \left(  \vec{\nabla} \dot{Z}_6 
- \frac{\tilde{Q} V \mu_6}{g_sk} L^2 \vec{u} \right) = 0 \ .
\label{eq:N_eqn}
\end{eqnarray}
If we define
\begin{equation}
\label{eq:K_eqn}
\vec{K} = c \vec{\nabla}^{-2} (\tilde{Q} \vec{u})\ ,
\end{equation}
for some constant $c$, then
\begin{equation}
\label{eq:curlcurlK1}
\vec{\nabla} \times ( \vec{\nabla} \times \vec{K}) = -\frac{cV} 
{(2\pi)^5 \alpha'^{5/2}g_s} \vec{\nabla} \dot{Z}_2
- c \tilde{Q} \vec{u} \ ,
\end{equation}
where we have used the expression for $Z_2$ in
equation~\eqref{eq:Z6_soln} and also charge conservation in the form
\begin{equation}
\partial_0 (\tilde{Q}) + \vec{\nabla} \centerdot (\tilde{Q}\vec{u}) =
0\ .
\end{equation}
Comparing~\eqref{eq:curlcurlK1} to the equation of
motion~\eqref{eq:A_eqn}, we find that with $c = {\mu_2 L^2}/{g_sk}$ we
get:
\begin{equation}
\label{eq:curlcurlK2}
\vec{\nabla} \times ( \vec{\nabla} \times \vec{K}) = -\vec{\nabla} \dot{Z}_2
- \frac{\tilde{Q} \mu_2}{g_sk} \vec{u} L^2\ .
\end{equation}
Similarly we can define
\begin{equation}
\label{eq:Ktilde_eqn}
\vec{\tilde{K}} = -\tilde{c} \vec{\nabla}^{-2} ( \tilde{Q} \vec{u})\ ,
\end{equation}
with $\tilde{c} = {V\mu_6L^2}/{g_sk}$, giving
\begin{equation}
\label{eq:curlcurlK_tilde}
\vec{\nabla} \times ( \vec{\nabla} \times \vec{\tilde{K}}) = -\vec{\nabla} 
\dot{Z}_6 + \frac{V \tilde{Q} \mu_6}{g_sk} \vec{u} L^2\ .
\end{equation}

Taking linear combinations of the equations of motion
\eqref{eq:A_eqn}, \eqref{eq:A_tilde_eqn} and~\eqref{eq:N_eqn}, and
using equations~\eqref{eq:curlcurlK2} and~\eqref{eq:curlcurlK_tilde}
we get
\begin{eqnarray}
\label{eq:eqn_motion1}
&&\hskip-1cm\vec{\nabla} \times \left( \frac{\vec{\nabla} \times (\vec{A} + Z_2^{-1/2}
Z_6^{1/2} \vec{N})} {Z_2^{-1} Z_6} 
 - \frac{\vec{\nabla} \times (Z_2^{1/2} Z_6^{1/2} \vec{N})} {Z_6}
\right)  =  \vec{\nabla} \times ( \vec{\nabla} \times \vec{K})\ ,
 \\  
\label{eq:eqn_motion2}
&&\hskip-1cm\vec{\nabla} \times \left( \frac{\vec{\nabla} \times (\vec{\tilde{A}} + 
Z_2^{1/2} Z_6^{-1/2} \vec{N})} {Z_2 Z_6^{-1}} 
 - \frac{\vec{\nabla} \times (Z_2^{1/2} Z_6^{1/2} \vec{N})} {Z_2}
\right)  =  \vec{\nabla} \times ( \vec{\nabla} \times
\vec{\tilde{K}})\ ,
\\
\label{eq:eqn_motion3}
&&\hskip-1cm\vec{\nabla} \times \left( \frac{\vec{\nabla} \times (\vec{A} + Z_2^{-1/2}
Z_6^{1/2} \vec{N})} {Z_6} 
+ \frac{\vec{\nabla} \times (\vec{\tilde{A}} + Z_2^{1/2} Z_6^{-1/2} \vec{N})}
{Z_2}
 - \frac{\vec{\nabla} \times (Z_2^{1/2} Z_6^{1/2} \vec{N})} {Z_2 Z_6}
\right)  =  0\ .
\end{eqnarray}

\subsection{The Effective Action}
\label{sec:effective-action}
We can integrate the equations of motion \eqref{eq:eqn_motion1} - 
\eqref{eq:eqn_motion3} to get
\begin{eqnarray}
&& \frac{\vec{\nabla} \times (\vec{A} + Z_2^{-1/2}
Z_6^{1/2} \vec{N})} {Z_2^{-1} Z_6} 
 - \frac{\vec{\nabla} \times (Z_2^{1/2} Z_6^{1/2} \vec{N})} {Z_6}
  =    \vec{\nabla} \times \vec{K} + \vec{\nabla} \alpha\ ,
\nonumber  \\  
&& \frac{\vec{\nabla} \times (\vec{\tilde{A}} + 
Z_2^{1/2} Z_6^{-1/2} \vec{N})} {Z_2 Z_6^{-1}} 
 - \frac{\vec{\nabla} \times (Z_2^{1/2} Z_6^{1/2} \vec{N})} {Z_2}
 =   \vec{\nabla} \times \vec{\tilde{K}} + \vec{\nabla}
 \tilde{\alpha}\ ,
\nonumber \\
\label{eq:int_eqn_motion3}
&& \frac{\vec{\nabla} \times (\vec{A} + Z_2^{-1/2} Z_6^{1/2} \vec{N})} {Z_6} 
+ \frac{\vec{\nabla} \times (\vec{\tilde{A}} + Z_2^{1/2} Z_6^{-1/2} \vec{N})} {Z_2}
 - \frac{\vec{\nabla} \times (Z_2^{1/2} Z_6^{1/2} \vec{N})} {Z_2 Z_6}
  =  \vec{\nabla} \nu\ ,
\end{eqnarray}
where $\nu$, $\alpha$ and $\tilde{\alpha}$ are functions of
integration. Taking divergences of the
equations~\eqref{eq:int_eqn_motion3}, we can show that it is
consistent to set $\alpha =\tilde{ \alpha} = \nu = 0$. (See \eg\ 
ref.\cite{Ferrell:thesis} for a discussion.) Then linear combinations
of equations~\eqref{eq:int_eqn_motion3} give:
\begin{eqnarray}
\vec{\nabla} \times (\vec{A} + Z_2^{-1/2} Z_6^{1/2} \vec{N}) & = & -
\vec{\nabla} \times \vec{\tilde{K}} \nonumber \\
\vec{\nabla} \times (\vec{\tilde{A}} + Z_2^{1/2} Z_6^{-1/2} \vec{N})&
= & - \vec{\nabla} \times \vec{K}  \nonumber \\
\label{eq:int_eqn_motion6}
\vec{\nabla} \times (Z_2^{1/2} Z_6^{1/2} \vec{N}) & = & -Z_2 \vec{\nabla} 
\times \vec{\tilde{K}} - Z_6 \vec{\nabla} \times \vec{K} \ .
\end{eqnarray}

Substituting equations~\eqref{eq:int_eqn_motion6} into the action
\eqref{eq:approx_action1} and integrating by parts gives:
\begin{eqnarray}
S^{\rm approx} & = & \int\! d^4\! x \,\, \bigg[  k  \bigg\{ - \dot{Z}_2 \dot{Z}_6 
+ \frac{1}{2} (\vec{\nabla} \times (\vec{A} + Z_2^{-1/2} Z_6^{1/2} \vec{N})) 
\centerdot (\vec{\nabla} \times \vec{K}) \nonumber \\
&&\hskip3cm+ \frac{1}{2} (\vec{\nabla} \times(\vec{\tilde{A}} + Z_2^{1/2} Z_6^{-1/2} 
\vec{N})) \centerdot (\vec{\nabla} \times \vec{\tilde{K}}) \bigg\} 
\nonumber \\
\label{eq:S_approx}
&&\hskip3cm+ L^2 \bigg\{ -\frac{\tilde{Q}}{g_s}\left(\mu_6 V - \mu_2 \right)
+ \frac{1}{2} \frac{\tilde{Q}}{g_s} \left( \mu_6 V Z_2 - \mu_2 Z_6 \right) u^2
\bigg\} \bigg]\ .
\end{eqnarray}
We now take the point--like limit, in which the charge density is
written as in equation~(\ref{eq:limit}). Then $Z_2$ and $Z_6$ are
given by the equations~\eqref{eq:Z6_soln}. Also the equations
\eqref{eq:K_eqn} and \eqref{eq:Ktilde_eqn} for $\vec{K}$ and
$\vec{\tilde{K}}$ have the solutions
\begin{eqnarray}
\vec{K} & = & -\frac{1}{4\pi} \frac{\mu_2 L^2}{g_sk} \sum_a \frac{Q_a}{r_a}
\vec{u}_a\ ,\quad {\rm and}\quad
\vec{\tilde{K}}  =  -\frac{1}{4\pi} \frac{V \mu_6 L^2}{g_sk} \sum_a 
\frac{Q_a}{r_a} \vec{u}_a \ ,
\end{eqnarray}
where $\vec{r}_a = \vec{x} - \vec{x}_a$. 

Then the first term in the action \eqref{eq:S_approx} becomes:
\begin{equation}
\label{eq:first_term}
\dot{Z}_2 \dot{Z}_6 = \sum_{a,b} \frac{(r_2)_a (r_6)_b}{r_a^3 r_b^3} \bigg\{
(\vec{r}_a \centerdot \vec{u}_a)  (\vec{r}_b \centerdot \vec{u}_b)
\bigg\}\ ,
\end{equation}
and the second term:
\begin{equation}
\label{eq:second_term}
(\vec{\nabla} \times \vec{K}) \centerdot (\vec{\nabla} \times \vec{\tilde{K}})
 = -\frac{1}{(4\pi)^2} \frac{V \mu_2 \mu_6 L^4}{g_s^2 k^2} \sum_{a,b}
 \frac{Q_aQ_b}{r_a^3r_b^3} \bigg\{ (\vec{r}_a \centerdot \vec{r}_b) (\vec{u}_a
 \centerdot \vec{u}_b) - (\vec{r}_a \centerdot \vec{u}_b) (\vec{r}_b
 \centerdot \vec{u}_a) \bigg\}\ .
\end{equation}
Consider the fifth term in the action \eqref{eq:S_approx}. Writing the
delta function in $\tilde{Q}$ of equation (\ref{eq:limit}) as
\begin{equation}
\delta^{(3)} (\vec{x} - \vec{x}_a) = \frac{1}{4\pi}\vec{\nabla}^2 \left( \frac{1}{r_a}
\right)\ ,
\end{equation}
then integrating by parts, we find
\begin{equation}
\label{eq:fourth_term}
\int\! d^3\! x 
\frac{L^2 \tilde{Q} u^2}{2g_s} \left( V \mu_6 Z_2 - \mu_2 Z_6 
\right)
= \sum_a \frac{L^2  Q_a u_a^2}{2g_s} \left( V \mu_6 - \mu_2 \right) - \frac{1}{4\pi}
\sum_{a,b} \int\! d^3\!x \frac{L^2 u_a^2}{2 \alpha' (2 \pi)^2} \frac{Q_aQ_b} 
{r_a^3r_b^3} (\vec{r}_a \centerdot \vec{r}_b)\ .
\end{equation}
Substituting equations~\eqref{eq:first_term}, \eqref{eq:second_term}
and \eqref{eq:fourth_term} into the action \eqref{eq:S_approx}, and
rearranging, and  defining:
\begin{equation}
S_{\rm eff} = \int\! L_{\rm eff} \,\, dt\ ,
\end{equation}
we find that (using equation~(\ref{eq:constantk}) for $k$): 
\begin{eqnarray}
L_{\rm eff} & = & -\frac{L^2}{g_s} \left(\mu_6 V - \mu_2 \right) \sum_a Q_a +
\frac{L^2}{g_s} \left(\mu_6 V - \mu_2 \right) \sum_a \frac{Q_a u_a^2}{2} 
\nonumber \\
\label{eq:L_eff}
&&+ \frac{L^2}{4 (2\pi)^3 \alpha'} \int\! d^3\!x \sum_{a,b}  \frac{Q_aQ_b} 
{r_a^3r_b^3} \left\{ (\vec{r}_a \times \vec{r}_b) \centerdot 
(\vec{u}_a \times \vec{u}_b) - \frac{1}{2} |\vec{u}_a - \vec{u}_b|^2 (\vec{r}_a
\centerdot \vec{r}_b) \right\} \ .
\end{eqnarray}

\subsection{Extracting the Metric}
\label{sec:two-body-metric}
For two wrapped branes, one of charge $Q_2$ and the other of charge
$Q_1$, equation~\eqref{eq:L_eff} reduces to
\begin{eqnarray}
L_{\rm eff} & = & -\frac{L^2}{g_s} \left(\mu_6 V - \mu_2 \right) (Q_1+Q_2) +
\frac{L^2}{g_s} \left(\mu_6 V - \mu_2 \right) \left( \frac{Q_1 u_1^2}{2} +
\frac{Q_2 u_2^2}{2} \right) 
\nonumber \\
\label{eq:L_eff1}
&&+ \frac{L^2}{4 (2\pi)^3 \alpha'} \int\! d^3\! x \,\, \frac{Q_1Q_2} 
{r_1^3r_2^3} \left\{ (\vec{r}_1 \times \vec{r}_2) \centerdot 
(\vec{u}_1 \times \vec{u}_2) - \frac{1}{2} |\vec{u}_1 - \vec{u}_2|^2 (\vec{r}_1
\centerdot \vec{r}_2) \right\} \ ,
\end{eqnarray}
where $r_1 = |\vec{x} - \vec{x}_1|$, and $r_2 = |\vec{x} - \vec{x}_2|$.

We can unpack this expression considerably. Consider the integral
\begin{equation}
\label{eq:I1}
I = \int\! d^3\! x \,\, \frac{(\vec{r}_1 \centerdot \vec{r}_2)}{r_1^3r_2^3}\ .
\end{equation}
We can introduce a Feynman parameter $\omega$ using the formula
\begin{equation}
\frac{1}{A^{\alpha}B^{\beta}} = \int_0^1\! d\omega \,\, \frac{\omega^{\alpha-1} 
(1-\omega)^{\beta-1}} {[\omega A + (1-\omega) B]^{\alpha+\beta}}
\frac{\Gamma(\alpha+\beta)}{\Gamma(\alpha) \Gamma(\beta)}\ .
\end{equation}
Then \eqref{eq:I1} becomes
\begin{equation}
\label{eq:I2}
I = \int\! d^3\! x \int_0^1\! d\omega \,\, \frac {\omega^{1/2} (1-\omega)^{1/2} (\vec{r}_1
\centerdot \vec{r}_2)} { [ \omega (x^2 - 2 \vec{x} \centerdot \vec{x}_1 +x_1^2)
+ (1-\omega) (x^2 - 2 \vec{x} \centerdot \vec{x}_2 + x_2^2)]^3}
\,\, \frac{\Gamma(3)}{\Gamma(\frac{3}{2})\Gamma(\frac{3}{2})}\ .
\end{equation}
Completing the square in the denominator in \eqref{eq:I2}, and substituting
$\vec{y} = \vec{x} - \omega \vec{x}_1 - (1-\omega) \vec{x}_2$ gives
\begin{equation}
I = \frac{\Gamma(3)}{\Gamma(\frac{3}{2})\Gamma(\frac{3}{2})}
\int_0^1\! d\omega \,\, \omega^{1/2} (1-\omega)^{1/2}
\int\! d^3y \,\, \frac{y^2 + (2\omega-1) \vec{y} \centerdot (\vec{x}_1 - \vec{x}_2)
 - \omega (1-\omega) (\vec{x}_1 - \vec{x}_2)^2} 
{ [y^2 + \omega (1-\omega) (\vec{x}_1 - \vec{x}_2)^2]^3}\ .
\end{equation}
Now 
\begin{equation}
\int\! d^3y \,\, \frac{(2\omega-1) \vec{y} \centerdot (\vec{x}_1 - \vec{x}_2)}
{[y^2 + \omega (1-\omega) (\vec{x}_1 - \vec{x}_2)^2]^3} = 0\ ,
\end{equation}
since the integrand is the sum of odd functions of the $y_i$. Therefore we can
write
\begin{equation}
I = \frac{\Gamma(3)}{\Gamma(\frac{3}{2})\Gamma(\frac{3}{2})} \int_0^1\! d\omega \,\,
\omega^{1/2} (1-\omega)^{1/2} \int\! d\Omega_2 \, dy \,\,
\frac{y^2(y^2-a^2)}{(y^2+a^2)^3}\ ,
\end{equation}
where $a^2 = \omega(1-\omega)(\vec{x}_1 - \vec{x}_2)^2 > 0$. We can do the $y$
integral using contour integration to get
\begin{eqnarray}
I & = & \frac{\Gamma(3)}{\Gamma(\frac{3}{2})\Gamma(\frac{3}{2})} \int_0^1\! 
d\omega \int\! d\Omega_2 \,\, \frac{\pi}{8|\vec{x}_1 - \vec{x}_2|}
 =  \frac{4\pi}{|\vec{x}_1 - \vec{x}_2|}\ . \label{eq:int1}
\end{eqnarray}
Using Feynman parameters again we can show that
\begin{equation}
\label{eq:int2}
\int\! d^3\! x \,\, \frac{(\vec{r}_1 \times \vec{r}_2) \centerdot 
(\vec{u}_1 \times \vec{u}_2)}{r_1^3 r_2^3} = 0\ .
\end{equation}

Substituting \eqref{eq:int1} and \eqref{eq:int2} into
\eqref{eq:L_eff1}, and setting $Q_1=Q_2=1$, we find that we can write the
metric as:
\begin{equation}
\label{eq:L_eff2}
L_{\rm eff} =
\left(C-\frac{D}{|\vec{x}_1-\vec{x}_2|}\right)(\vec{u}_1^2+\vec{u}_2^2)+\frac{2D}{|\vec{x}_1-\vec{x}_2|}\vec{u}_1\cdot\vec{u}_2\ ,
\end{equation}
where
\begin{equation}
C=\frac{2L^2}{g_s}(\mu_6V-\mu_2)\ ,\qquad {\rm and}
\qquad D=\frac{2L^2}{(4\pi)^2\alpha^\prime}\ .
  \label{eq:CandD}
\end{equation}
The two terms in \eqref{eq:L_eff2} controlled by the constant $C$
describe the motion of the centre of mass moduli, for which the metric
on moduli space is flat.  However, the terms controlled by $D$
determine the relative motion of the branes. Writing everything in
terms of the relative coordinates, $r=|\vec{x}_1-\vec{x}_2|,
\vec{u}=d\vec{r}/dt$, we can write:
\begin{equation}
ds^2 = \bigg\{1 - \frac{\ell_{\rm e}}{r} \bigg\} (dr^2 + r^2 d\Omega^2_2)\
,\qquad \ell_{\rm e}=\frac{D}{C}= \frac{\alpha^\prime e^2}{4}\ ,\qquad \frac{1}{e^2}=\frac{{\alpha^\prime}^{1/2}}{g_s}\left(\frac{V}{V^*}-1\right)\ ,
  \label{eq:relative}
\end{equation}
where we have used standard coordinates $(r,\theta,\phi)$ on $\IR^3$,
and $d\Omega_2^2$ is the metric on a round two--sphere, with
coordinates $(\theta,\phi)$. Note the satisfying relation between
$1/e^2$ and the enhan\c con radius for $N=1$, from
equation~(\ref{enhanconradius}).

In fact, the result in equation~(\ref{eq:L_eff2}) is the essential
content of the computation so far. The full result in
equation~(\ref{eq:L_eff}) is really only a sum of two--body
interactions. In view of the work of Shiraishi for the ``$a=1$''
Einstein--Maxwell--Dilaton system\cite{Shiraishi:1993ia}, this is to
be expected\footnote{For more on the many--body interpretation of the
  interaction terms for various types of dilaton black holes, see
  ref.\cite{Gibbons:1997iy}.}. The structure of our computations is in
that class. So our result for an arbitrary number of branes can be
obtained by summing over the result for the two body case and so we
can write the metric on moduli space in the following form:
\begin{eqnarray}
ds^2&=&g_{ab}d\vec{x}^ad\vec{x}^b\ ,\qquad{\rm where}\nonumber\\
g_{ab}&=&\frac{\ell_{\rm e}}{|\vec{x}_a-\vec{x}_b|}\ ,\quad a\neq b\ ,\nonumber\\
g_{aa}&=&1-\sum_{b\neq a}\frac{\ell_{\rm e}}{|\vec{x}_a-\vec{x}_b|}\ ,\qquad\mbox
{\rm  no sum on $a$}\ .
\label{themetric}
\end{eqnarray}

This is not quite the final result. Throughout, we have neglected $N$
parameters of the solution, one for each of the wrapped D--branes. We
should take these into account, since there is non--trivial structure
on their moduli space. These parameters are, in the monopole language,
the canonical conjugates to the electric charges that each monopole
can be endowed with by a duality transformation, as originally shown
by Julia and Zee in ref.\cite{Julia:1975ff}. There is also an
excellent discussion of these parameters by Zee in
ref.\cite{Zee:2003mt}, and by Gibbons and Manton in
ref.\cite{Gibbons:1986df}.  The point is that these parameters appear
as natural {\it physical} phases of the monopole solution. It is
interesting to see how these extra degrees of freedom are included in
our computation here.

\subsection{The  Phases}
\label{sec:relative-phases-1}
In the wrapped D6--brane language, there is an elegant description of
the extra parameters (or internal phases) corresponding to the
electric charges of the monopoles.  The key piece of the construction
was presented in ref.\cite{Johnson:1999qt}, but here we will need to
augment it considerably. To recapitulate, the D6--brane part of the
solution contributes a Dirac magnetic monopole source, which we will
call $C^{(1)}_{\rm D}$ which comes from Hodge--dualising $C^{(7)}$ in
ten dimensions. In the world--volume action of the brane,
$C^{(1)}_{\rm D}$ couples, through its pull--back, to the
world--volume gauge field strength $F_{\alpha\gamma}={\bf t}^a
F^a_{\alpha\gamma}$ as
follows\cite{Douglas:1998uz,Li:1996pq,Witten:1996im}:
\begin{equation}
-2\pi \alpha' \mu_2 \int \Tr\left[C^{(1)}_{\rm D} \wedge F\right]\ .
\label{Cinteraction}
\end{equation} 
Of course, $F$ is the field strength for a gauge group as large as
$U(N)$, but in the present context of separated branes, we are working
in the Abelian case where only the maximal $U(1)^N$ subgroup survives.
In the non--Abelian case, we would also have to include contributions
arising from the self--interactions of the (adjoint) scalar fields
corresponding to the positions of the individual wrapped branes, as in
ref.\cite{Myers:1999ps}. The field $C^{(1)}_{\rm D}$ itself would be sensitive
to this non--Abelian physics, since it depends on the brane positions.
In the Abelian limit, however, things are much simpler, {\it at least
in the limit of extreme monopole separations \ie\ , the classical limit
in the dual $U(N)$ gauge theory}. There is no interaction between the
different $U(1)$s. A natural basis for the generators ${\bf t}^a$ in
the $N{\times}N$ fundamental representation is:
\begin{equation}
  \label{eq:generators}
{\bf t}^a={\rm diag}\{\ldots 0,\ldots,1,\ldots,0,\ldots\}\ ,
\end{equation}
where there is a single entry of unity in the $a$th position along the
diagonal.  The trace in the above equation~\eqref{Cinteraction} will
give zero for all of the $U(1)$ generators except the diagonal one,
and so we end up with a coupling only to the diagonal $\sum_{a=1}^N
F^a_{\alpha\gamma}$.

This is not the only modification, however. In the Dirac--Born--Infeld
action we treated earlier in section~\ref{sec:pert-stat-solut}, we did
not include the coupling to the gauge fields $F^a_{\alpha\gamma}$. The
modification is (again, restricting to the Abelian situation and
wrapping on K3):
\begin{equation}
S_{\rm matter}=-\int\! d^3\!\xi\,\, e^{-\Phi}(T_6Ve^{\beta}-T_2)\,\,\Tr\left[-{\rm
    det}(G_{\alpha\gamma}+2\pi\alpha^\prime {\bf t}^aF^a_{\alpha\gamma}) \right]^{1\over2}\ ,
  \label{eq:mattermodified}
\end{equation}
and since we can write 
\begin{equation}
-{\rm
    det}(G_{\alpha\gamma}+2\pi\alpha^\prime {\bf t}^aF^a_{\alpha\gamma}) = \left(-{\rm det}
    G\right)\left(1+\frac{1}{2}{\bf t}^a{\bf t}^b
G^{\alpha\delta}G^{\gamma\epsilon}F^a_{\alpha\gamma}F^b_{\delta\epsilon}\right)\ ,
  \label{eq:expand}  
\end{equation}
and recalling that $\Tr\left( {\bf t}^a{\bf t}^b\right)=\delta_{ab}$,
we see that the trace gives us $N$ independent contributions from each
of the $N$ gauge fields.

The point is that in this three--dimensional world--volume, the $a$th
gauge field $A_\alpha^a$, can be dualised to give a scalar, $s^a$.
These $N$ scalars are the phases of the monopoles in the reduced
action. We can write an equivalent action for them by following a
slight generalisation of ref.\cite{Schmidhuber:1996fy}, introducing a
family of auxiliary fields $f^a_\alpha$, and writing a slightly
different action. The term $2\pi\alpha^\prime F^a_{\alpha\gamma}$ in
the Dirac--Born--Infeld action is replaced by
$e^{2\Phi}(\tau_6Ve^{\beta}-\tau_2)^{-2}f^a_\alpha f^a_\gamma$ (no sum
on $a$ here), and the terms $\sum_a F^a\wedge f^a$ are added to the
Lagrangian. Integrating out the $f^a$ will give us the action we had
previously, while integrating out the potential $A^a_\alpha$ instead
will give us the equations:
\begin{equation}
\epsilon^{\alpha\gamma\delta}\partial_\gamma((C^{(1)}_{\rm D})_\delta-f^a_\delta)=0\ ,
  \label{eq:enforce}
\end{equation}
where the world--volume index on $C^{(1)}_{\rm D}$ arises from the
pull--back, \ie\ absorbing the spacetime index with a factor $\partial
x^\mu/\partial \xi^\delta$, where $\xi^\delta$ are world--volume coordinates.
The solution of the equation~(\ref{eq:enforce}) defines a family of
scalars, $s^a$:
\begin{equation}
\partial_\alpha s^a=(C^{(1)}_{\rm D})_\alpha-f^a_\alpha\ .
  \label{eq:scalars}
\end{equation}
We can now eliminate $F^a_{\alpha\gamma}$ from our action and substitute  for
the $f^a_\alpha$ using equation~(\ref{eq:scalars}), with result:
\begin{eqnarray}
S_{\rm matter}&=&-\int\! d^3\!\xi \,
e^{-\Phi}(T_6Ve^{\beta}-T_2)\times\nonumber\\
&&\hskip0.1cm \Tr\left[-{\rm
    det}\{G_{\alpha\gamma}+e^{2\Phi}(\tau_6Ve^{\beta}-\tau_2)^{-2}{\bf t}^a(\partial_\alpha s^a-(C^{(1)}_{\rm D})_\alpha)(\partial_\gamma s^a-(C^{(1)}_{\rm D})_\gamma)\} \right]^{1\over2}.
  \label{eq:complete}
\end{eqnarray}
To compute $C^{(1)}_{\rm D}$, we must use its definition {\it via} ten
dimensional Hodge duality:
\begin{equation}
d C^{(1)}_{\rm D}=e^{-\frac32\Phi}\,{}^*\!F^{(8)}=
e^{-\frac32\Phi}\,{}^*\!dC^{(7)}\ ,
  \label{eq:dualdefinition}
\end{equation}
which gives, after some algebra that $C^{(1)}_{\rm D}$ is given by:
\begin{equation}
\vec{\nabla}\left( Z_6\right)=\vec{\nabla}\times C^{(1)}_{\rm D}\ .
  \label{eq:dirac}
\end{equation}
For example, if we  use $Z_6$ as given in equation~(\ref{eq:Z6_soln}), and
we choose to use coordinates $(r,\theta,\phi)$ on the
three--dimensions transverse to the branes, then
\begin{eqnarray}
C^{(1)}_{\rm D} & = & r^2 Z_6^\prime \cos\theta
\,d\phi=-\sum_a\left\{\frac{(r_6)_a}{|\vec{r}-\vec{r}_a|^3}r(r^2-\vec{r}\cdot\vec{r}_a)\right\}\cos\theta
d\phi\ ,
\end{eqnarray}
which for $\vec{r}_a=0$, reduces to the familiar charge $N$ monopole
potential:
\begin{equation}
C^{(1)}_{\rm D}=-r_6 N \cos\theta\, d\phi\ .
  \label{eq:simpleDirac}
\end{equation}
In principle, we must redo the computations we carried out in the
previous sections, inserting the perturbed fields into the modified
``matter'' and ``current'' actions that we have written
above. However, it is important to note that the method will miss some
crucial parts of the computation unless we augment the procedure
somewhat. The crucial point is that the supergravity techniques that
we have used so far are insensitive to the $N$ different $U(1)$s
which reside on the D--branes. They are only sensitive to the overall
diagonal $U(1)$ which refers to the center of mass of the system.

In particular this means that the system will not be able to generate
terms which couple different $U(1)$s together, which is a non--trivial
feature of the moduli space, arising from terms such as the
off--diagonal $g_{ab}$ derived in the previous section.  From the
point of view of the world--volume $U(N)$ 2+1 dimensional gauge
theory, such coupling corresponds to important one--loop corrections
to the classical decoupled result for the Coulomb branch moduli
space\cite{Seiberg:1996nz,Chalmers:1997xh,Hanany:1997ie}. As may have
been observed in the previous sections, the crucial tool which
generates such terms is the $\delta$--function regulator (discussed in
section~\ref{sec:point-sourc-regul}) which converts the smeared
charge/mass distribution to the point--like ones at the end of the
computation. In section~\ref{sec:effective-action}, it is this
technique that extracts the terms that allow the $a$th centre to
interact with the $b$th one. We now need this to be correlated with
the individual $U(1)$s we are studying in this section, correctly
tying the $a$th $U(1)$ (and hence the phase $s^a$) with the $a$th
position $\vec{x}^a$. For the regulator tool to be sensitive to the
different $U(1)$s, it needs to be modified to something like:
\begin{equation}
{\tilde Q}=\sum_a {\bf t}^a\delta^{(3)}(x-x_a)\ ,
  \label{eq:regulator}
\end{equation}
where it is to be understood this operator is to be used under the
gauge trace in the actions written here. The facility of this
modification is that on objects which are charged under the diagonal
$U(1)$ it will count the total R--R charge in the usual way, since to
be diagonally charged is to carry the identity matrix as the
generator.  Everything from the previous section falls into this
category.  However, our object is now more refined, since it can also
be sensitive to the individual $U(1)$s. Our supergravity computations
now have a chance of uncovering the subtle terms.

Having noted the shortcomings of the direct supergravity computation,
we can proceed more rapidly as follows. The brane labeled ``$a$'' has
coordinate $\vec{x}^a$, and we have already computed the metric
$g_{ab}$ on these coordinates in the last section. This result clearly
tells us what the metric for the kinetic terms of the $U(1)$s is
directly, telling us precisely how the $a$th $U(1)$ is coupled. The
coordinates $\vec{x}^a$ are simply the adjoint scalars in the $a$th
$U(1)$ gauge  supermultiplet. So the output of expanding
out the $U(1)^N$ sector to leading order (quadratic in
$F_{\alpha\gamma}$) is easy to write:
\begin{equation}
S_{\rm YM}=\int\! d^3\xi
\left(
  -\frac{1}{4e^2}g_{ab}F^a_{\alpha\gamma}F^{b\alpha\gamma}\right)\
,
  \label{eq:yangMills}
\end{equation}
where $g_{ab}$ is given in equations~(\ref{themetric}), and we've used
the standard value of the bare Yang--Mills coupling on the brane,
taking into account the wrapping on a K3 of asymptotic volume $V$. The
constant part of $g_{ab}$ (for $a{=}b$), which is $C$ given in
equation~(\ref{eq:CandD}), is the classical result giving the basic
coupling of each individual $U(1)$, while the rest of the metric is
one--loop, from the point of view of the three--dimensional $U(N)$
gauge theory\cite{Seiberg:1996nz,Chalmers:1997xh,Hanany:1997ie}.

There is another set of terms that we can deduce. These terms again
follow in principle from the fact that we know that the $U(1)$ on each
brane must couple to the pull--back of the dual of the R--R scalar
produced by all the other branes. This is again something that the
supergravity expressions need some encouragement to produce, since
they are naturally adapted to producing only the diagonal $U(1)$ data.
With care, however, we can see how it must work. The $a$th brane has a
Chern--Simons coupling of the gauge field $\sum_a{\bf
  t}^aF^a_{\alpha\gamma}$ to the Dirac monopole R--R field produced by
the brane at $\vec{x}^b$. In the supergravity computation, that field
descends from $C^{(1)}_{\rm D}$, whose curl is given by the gradient
of $|\vec{x}-\vec{x}_a|^{-1}$, as we have discussed above. In the
computation, a regularising $\delta$--function will produce a
$|\vec{x}_b-\vec{x}_a|^{-1}$ from this. Morally speaking, the
supergravity field should contain the information that there is a
generator ${\bf t}^b$ associated to this term as well. This survives
in the remnant, denoted $\vec{\omega}_{ab}$, of $C^{(1)}_{\rm D}$ which is
the field at $a$ due to $b$. Its curl is, up to a numerical factor,
the gradient of $|\vec{x}_b-\vec{x}_a|^{-1}$, and the dependence on
the generator ${\bf t}^b$ picks out the component
$F^{b}_{\alpha\gamma}$ on $a$'s world--volume after taking the trace.
We must pull the spacetime index of the vector $\vec{\omega}_{ab}$
back to a world--volume index using the coordinate on the $a$th
world--volume, and we do this by forming the dot product with
$\partial\vec{x}^a/\partial\xi^\gamma$.  In summary, the term is
simply:
\begin{equation}
S_{\rm FRR}=-\frac{1}{8\pi}\int\! d^3\!\xi\,\,\epsilon^{\alpha\gamma\kappa}F^b_{\alpha\gamma}\vec{\omega}_{ab}
\cdot\partial_\kappa\vec{x}^a\ .
  \label{eq:chernsimons}
\end{equation}

Finally, in three dimensions we can introduce the analogue of four
dimensional $\theta$--angles for each gauge group, since there is a
natural topological invariant measuring the analogue of instanton
number (winding at infinity) in the gauge
field\cite{Polyakov:1977fu,Zee:2003mt}:
\begin{equation}
S_{\rm wind}=\frac{1}{8\pi}\int\! d^3\!\xi\,\,\epsilon^{\alpha\gamma\kappa}\partial_\kappa F^a_{\alpha\gamma}s_a\ ,
  \label{eq:instanton}
\end{equation}
where the $s^a$ are $2\pi$ periodic, since $e^{iS_{\rm
    wind}}=e^{in^as_a}$ where $n^a$ is an integer denoting the amount
of ``instanton number'' in the $a$th $U(1)$.  Clearly, if we treat the
$s_a$ as dynamical variables in the full problem $S_{\rm total}=S_{\rm
  YM}+S_{\rm FRR}+S_{\rm wind}$, they are simply Lagrange multipliers
which set the~$n^a$ to zero. Alternatively, we can integrate out the
$F^a_{\alpha\beta}$ to which they
couple\cite{Townsend:1996af,Seiberg:1996nz}. Their equation of motion
is simply:
\begin{equation}
F^a_{\alpha\gamma}=-\frac{e^2}{4\pi}(g^{-1})^{ab}\epsilon_{\alpha\gamma}^{\phantom{\alpha\gamma}\kappa}
(\partial_\kappa s_b+\vec{\omega}_{bc}\cdot\partial_\kappa\vec{x}^c)\ .
  \label{eq:integrateF}
\end{equation}
Substituting this back into the action, we get the simple result:
\begin{equation}
S_{\rm phases}=\frac{1}{2}\int\! d^3\!\xi \, 
\frac{e^2}{(4\pi)^2}(g^{-1})^{ab}(\partial_\kappa
s_a+\vec{\omega}_{ac}\cdot\partial_\kappa\vec{x}^c)(\partial^\kappa
s_b+\vec{\omega}_{bc}\cdot\partial^\kappa\vec{x}^c)\ .
  \label{eq:phases}
\end{equation}

\subsection{The Final Metric}
\label{sec:final-metric}
So upon reducing the result~(\ref{eq:phases}) on $T^2$, the $s^a$ are
just functions of $t$, and including the result~(\ref{themetric}) from
the previous section, we see that our complete metric on the $4N$
dimensional moduli space of our $N$ wrapped D6--branes is:
\begin{equation}
ds^2=g_{ab}d\vec{x}^a\cdot
d\vec{x}^b+(g^{-1})^{ab}(ds_a+\vec{\omega}_{ac}\cdot
d\vec{x}^c)(ds_b+\vec{\omega}_{bc}\cdot d\vec{x}^c)\ ,
  \label{eq:finalmetric}
\end{equation}
where $g_{ab}$ is given in equation~(\ref{themetric}), and
$\vec{\nabla}\times\vec{\omega}_{ab}=\vec{\nabla}(g_{ab})$. This is
the Gibbons--Manton metric\cite{Gibbons:1995yw} for $N$
well--separated BPS monopoles, demonstrating that our K3--wrapped
D6--branes are indeed behaving as non--coincident BPS monopoles.

\section{Summary}
Our result is satisfying. We have confirmed the expectation that K3
wrapped D6--branes in type IIA string theory behave like BPS
monopoles\cite{Johnson:1999qt}, by showing that the metrics on moduli
spaces for $N$ objects exactly match, at least in perturbation theory.
We expect that exactly the same arguments that the non--perturbative
corrections complete the monopole moduli space into a unique 4$N$
dimensional hyper--K\"ahler manifold generalising the Atiyah--Hitchin
manifold will apply to our case as well. This also fits well with the
fact that this is isomorphic to the geometry of the Coulomb branch of
2+1 dimensional pure $U(N)$ gauge
theory\cite{Seiberg:1996nz,Chalmers:1997xh,Hanany:1997ie}. The
presence of this gauge theory is extremely natural in this picture. It
is simply the gauge theory on the world--volume of the K3--wrapped
D6--branes.

We also notice that we have essentially indirectly computed the same
result for $N$ of the objects made by binding
together\cite{Johnson:1996bf} a Kaluza--Klein monopole and an
H--monopole (wrapped NS5--brane) for the case of the heterotic string
compactified on $T^4$, since our basic action  has a simple symmetry
which performs the strong/weak coupling duality transformation between
the two string theories. These objects are confirmed as BPS monopoles
as well\cite{Johnson:2001wm,Krogh:1999qr}.

As mentioned in the introduction, the kinds of study we have presented
here are very interesting and worthwhile to explore in the detail that
we have carried out here. D--branes in flat spacetime are already
interesting and instructive, having fueled most of the discoveries
since (and been used to check the consistency of) the Second
Revolution. We know that they get much more structure when put into
more interesting situations, such as wrapping, intersecting, coupling
to background fields, \etc. We have learned much from these new
situations already, and further studies will certainly teach us about
even more new phenomena which will augment our understanding of the
physics of gauge theories, spacetime, geometry, and perhaps even more.

\section*{Acknowledgements}
JKB is supported by an EPSRC studentship. CVJ's research while at
Durham was supported in part by the EPSRC.  CVJ's research while at
the ITP was supported by the National Science Foundation under Grant
No.  PHY-99-07949.


\providecommand{\href}[2]{#2}\begingroup\raggedright\endgroup



\end{document}